\documentclass[a4paper,10pt,DIV=calc]{scrartcl}

\usepackage{cancel}
\usepackage{tabularx}
\usepackage{subfig}
\usepackage{a4}
\usepackage{amsmath,amssymb}
\usepackage{bbm}
\usepackage{amsthm}
\usepackage{dsfont}
\usepackage[onehalfspacing]{setspace} 
\usepackage{appendix}
\usepackage{graphicx}
\usepackage{enumerate}
\usepackage{hyphenat}
\usepackage{cite}
\usepackage{colonequals}
\usepackage{relsize}
\usepackage{caption}
\usepackage{setspace}
\usepackage{MnSymbol}

\usepackage{enumerate}
\usepackage{relsize}

\usepackage[usenames,dvipsnames]{xcolor}
\usepackage[all]{xypic}

\usepackage{accents}
\usepackage{amsmath,amssymb}
\makeatletter
\newsavebox\myboxA
\newsavebox\myboxB
\newlength\mylenA

\newcommand*\xo[2][0.75]{%
    \sbox{\myboxA}{$\m@th#2$}%
    \setbox\myboxB\null
    \ht\myboxB=\ht\myboxA%
    \dp\myboxB=\dp\myboxA%
    \wd\myboxB=#1\wd\myboxA
    \sbox\myboxB{$\m@th\overline{\copy\myboxB}$}
    \setlength\mylenA{\the\wd\myboxA}
    \addtolength\mylenA{-\the\wd\myboxB}%
    \ifdim\wd\myboxB<\wd\myboxA%
       \rlap{\hskip 0.5\mylenA\usebox\myboxB}{\usebox\myboxA}%
    \else
        \hskip -0.5\mylenA\rlap{\usebox\myboxA}{\hskip 0.5\mylenA\usebox\myboxB}%
    \fi}
\makeatother

\usepackage{bbm}
\usepackage{amsthm}
\usepackage{appendix}
\usepackage{graphicx}
\usepackage[para]{footmisc}
\usepackage{tablefootnote}

\usepackage[para]{threeparttable}
\usepackage{etoolbox}
\usepackage{lipsum}
\usepackage[all]{xypic}
\usepackage{color}
\AtEndEnvironment{threeparttable}{\vskip10pt}{}{}
\makeatletter
\patchcmd{\TPT@doparanotes}
  {\TPT@hsize}
  {\TPT@hsize\footnotesize}
  {}
  {}
\makeatother

\newcommand{\PT}{\ensuremath{\mathrm{P}}}
\newcommand{\GTp}{\ensuremath{\mathrm{GT'}}}
\newcommand{\GT}{\ensuremath{\mathrm{GT}}}
\newcommand{\KS}{\ensuremath{\mathrm{KS}}}
\newcommand{\NW}{\ensuremath{\mathrm{NW}}}
\newcommand{\no}[3]{\ensuremath{(#1\,#2\,#3)}}
\newcommand{\nono}[3]{\ensuremath{\lbrace#1\,#2\,#3\rbrace}}
\newcommand{\ve}[3]{\ensuremath{[#1\,#2\,#3]}}
\newcommand{\veve}[3]{\ensuremath{\langle#1\,#2\,#3\rangle}}

\newcommand{\bbb}{\ensuremath{\mathbf{b}}}
\newcommand{\vv}{\ensuremath{\mathbf{v}}}
\newcommand{\uu}{\ensuremath{\mathbf{u}}}
\newcommand{\mm}{\ensuremath{\mathbf{m}}}

\newcommand{\nn}{\ensuremath{\mathbf{n}}}
\newcommand{\ee}{\ensuremath{\mathbf{e}}}
\newcommand{\ff}{\ensuremath{\mathbf{f}}}
\newcommand{\al}{\ensuremath{\alpha}}
\newcommand{\be}{\ensuremath{\beta}}
\newcommand{\PP}{\mathcal{P}^{24}}

\newcommand{\cof}[1]{\operatorname{cof} #1\,}
\newcommand{\aand}{\mbox{ and }}

\newcommand{\dd }{\cdot}
\newcommand{\hl}{\rule[.5ex]{1.2em}{0.4pt}}
\newcommand{\sgn}{\operatorname*{sgn}}
\DeclareMathOperator{\tr}{Tr} 
\graphicspath {{figures/}}

\newcommand{\id}{{\mathbb{I}}}

\newcommand{\ssp}[2]{( #1 \dd #2 )}
\newcommand{\tp}{\otimes}
\newcommand{\quo}[1]{{``#1''}}
\newcommand{\alp}{_{\alpha'}}

\definecolor{mygray}{gray}{0.7}

  \usepackage{authblk}
\usepackage{fancyhdr}

\usepackage{hyperref}
\makeatletter
\def\@maketitle{%
  \newpage
  \null
  \vskip 2em%
  \begin{center}%
  \let \footnote \thanks
    {\Large\bf  \@title \par}%
    \vskip 1.5em%
    {\normalsize
      \lineskip .5em%
      \begin{tabular}[t]{c}%
        \@author
      \end{tabular}\par}%
    \vskip 1em%
    {\normalsize \@date}%
  \end{center}%
  \par
  \vskip 1.5em}
\makeatother
\title{A Theoretical Investigation of Orientation Relationships and Transformation Strains in Steels}

\author{Konstantinos Koumatos%
  \thanks{\texttt{konstantinos.koumatos@gssi.infn.it}}}
\affil{\small\textit{Gran Sasso Science Institute, }\\ \textit{Viale Fransesco Crispi 7,} \\ \textit{67100 L'Aquila, Italy}}

\author{Anton Muehlemann%
  \thanks{\texttt{muehlemann@maths.ox.ac.uk}}}
\affil{\small\textit{Mathematical Institute, University of Oxford,}\\\textit{ Andrew Wiles Building, Radcliffe Observatory Quarter, Woodstock Road,} \\ \textit{Oxford OX2 6GG, United Kingdom}}

\date{Dated: \today}

 \hypersetup{
  pdfauthor = {Konstantinos Koumatos (Gran Sasso Science Institute) and Anton Muehlemann (University of Oxford)},
  pdftitle = {A theoretical investigation of orientation relationships and transformation strains in steels},
  pdfsubject = {MSC (2010): 74A05, 74N05, 74N10},
  pdfkeywords = {Nishiyama-Wassermann, Kurdjumov-Sachs, tetragonal, orientation relationships, transformation strain, steel, fcc to bcc, fcc to bct, Pitsch, Greninger-Troiano, Bain, Inverse Greninger-Troiano, face-centred cubic, body-centred cubic, body-centred tetragonal}
}

\begin{document}

\maketitle
\begin{abstract}
The identification of orientation relationships (ORs) plays a crucial r\^ole in the understanding of solid phase transformations. In steels, the most common models of ORs are the ones by Nishiyama-Wassermann (\NW) and Kurdjumov-Sachs (\KS). The defining feature of these and other OR models is the matching of directions and planes in the parent face-centred cubic $\gamma$-phase to ones in the product body-centred cubic/tetragonal $\alpha/\alpha'$-phase.

In this paper a novel method that identifies transformation strains with ORs is introduced and used to develop a new strain-based approach to phase transformation models in steels. Using this approach, it is shown that the transformation strains that leave a close packed plane in the $\gamma$-phase and a close packed direction within that plane unrotated are precisely those giving rise to the \NW\ and \KS\ ORs when a cubic product phase is considered. Further, it is outlined how, by choosing different pairs of unrotated planes and directions, other common ORs such as the ones by Pitsch (\PT) and Greninger-Troiano (\GT) can be derived. 

One of the advantages of our approach is that it leads to a natural generalisation of the \NW, \KS\ and other ORs for different ratios of tetragonality $r$ of the product bct $\alpha'$-phase. These generalised ORs predict a sharpening of the transformation textures with increasing tetragonality and are thus in qualitative agreement with experiments on steels with varying alloy concentration.
 \vspace{4pt}



\vspace{4pt}
\end{abstract}
\tableofcontents

\section{Introduction}
The transformation mechanism from the face-centred cubic (fcc) to the body-centred cubic/tetragonal (bcc/bct) phase of steel has received widespread attention and the most influential early studies include \cite{Bain,KS,Nishiyama,Wassermann}. In his seminal paper, Bain \cite{Bain} proposed a mechanism that transforms the fcc $\gamma$-phase of iron to its bcc $\alpha$-phase \quo{requiring the least temporary distortion}. His conceived mechanism, although now widely accepted, was not without criticism from his contemporaries. Among the critics were Kurdjumov and Sachs \cite{KS} who conducted X-ray diffraction measurements on $1.4\%$ carbon steel and measured the orientation relationships between austenite and pure bcc $\alpha$-iron as well as between austenite and $1.4\%\, \mathrm C$ $\al'$-steel.\footnote{Henceforth, we adopt the convention from \cite{Nishiyamabook} of using the symbol ${\al'}$ for the low temperature phase of steels irrespectively of whether it is cubic or tetragonal.} The most important 
feature 
of their mechanism was the 
parallelism between the $\no{1}{1}{1}_\gamma$ and the $\no{0}{1}{1}_{\al'}$ plane as well as the $\ve{1}{0}{\bar 1}_\gamma$ and the $\ve {1}{\bar 1}{1}_{\al'}$ direction and they explained how these conditions can be satisfied by a combination of three shears. Following their construction step by step one sees that the overall deformation is always one of the Bain strains followed by a rigid body rotation and that the resulting orientation relationship for pure iron differs from the one for  $1.4\%\, \mathrm C$ steel (see Tables $2$ in \cite{KS} and \cite{Otte}). In 1934, using the same methods, Nishiyama \cite{Nishiyama} investigated a $\mathrm{Fe}$-$30\% \,\mathrm{Ni}$ single crystal which, like pure iron, undergoes an fcc to bcc transformation. Based on his observations, Nishiyama proposed a different orientation relationship that has the same parallel planes but the direction $\ve{1}{0}{\bar 1}_\gamma$ parallel to $\ve {1}{0}{0}_{\al'}$. One year later, Wassermann \cite{
Wassermann} independently postulated the same relationships and also confirmed the earlier results by Kurdjumov and Sachs. Apart from the Nishiyama-Wassermann (\NW) and Kurdjumov-Sachs (\KS) orientation relationships (ORs) several other ORs, e.g. by Pitsch \cite{Pitsch} (\PT) and Greninger-Troiano \cite{GT} (\GT), have been proposed and they all share the common feature of matching directions and planes in the parent phase to ones in the product phase.

In the present article, we would like to shift this paradigm towards a derivation of orientation relationships based on the transformation strains. Compared to previous approaches (see e.g. \cite{Guo,JonasMeteor,CayronActa}), our approach brings the following novelties:
\begin{enumerate}
 \item The only necessary inputs are the lattice parameters of the two phases and the knowledge of a plane and a direction that is left unrotated.
 \item Each derived strain can be uniquely idenfied with an OR and the parallelism between planes and directions in the two phases follows. 
 \item The additional knowledge of the actual underlying deformation of the material can e.g. be used to unambiguously determine twin relationships (cf. Section~\ref{SecTwin}) and generally lay the groundwork for mathematical theories of steels based on energy minimisation (see e.g. \cite{Bha,NewPersp}).
 \item Our method takes into account the ratio of tetragonality $r=c/a$ of the bct $\al'$ phase. Thus, the derived strains and orientation relationships also depend on $r$ and can be expressed explicitly as functions of $r$. 
  \end{enumerate}
 For $r=1$, corresponding to bcc, we recover the original \NW, \KS\ and \PT\ ORs. However, for $r>1$, our approach predicts a deviation from the original ORs. We show how this leads to a sharpening of the transformation textures and how it can be used to explain the deviation from the exact parallelism condition in the \GT\ ORs.

The structure of the paper is as follows: at the end of this section we clarify the notation that will be used throughout. In Section~\ref{SecUnif}, we introduce a unified approach for the derivation of phase transformation models in steels which entails a general method to identify transformation strains with orientation relationships. In Section~\ref{SecDer}, we apply our unified approach to deduce the \KS\ and \NW\ transformation strains and orientation relationships; we also comment on how the obtained ORs relate to other common descriptions of the \NW\ and \KS\ ORs and show how the additional knowledge of the strains can be used to unambiguously determine twin relationships between \KS\ variants. At the end of Section~\ref{SecDer}, we illustrate how according to our unified approach the \KS\ and \NW\ ORs change with increasing ratio of tetragonality $r$ of the $\al'$ phase. In Section~\ref{SecOtherORs}, we indicate how the same methods can be used to explain and generalise the Pitsch (\PT), Greninger-
Troiano (\GT) and inverse Greninger-Troiano (\GTp) OR models.

\subsection*{Preliminaries}
Let us consider an orthonormal basis $\{\ff_1,\ff_2,\ff_3\}$. By $\ve{a}{b}{c}=\tfrac{a \ff_{1}+ b \ff_{2}+ c \ff_{3}}{\sqrt{a^2+b^2+c^2}}$ we denote a normalised direction expressed in this basis.\footnote{As is commonly asserted in the literature, we make the identification $-a = \bar a$.} Similarly, by $\no{a}{b}{c}$ we denote a normal in the same basis.\footnote{Note that since  $\{\ff_1,\ff_2,\ff_3\}$ is an orthonormal basis it coincides with its reciprocal basis, i.e. \ve{a}{b}{c}=\no{a}{b}{c}.} For $\uu=\ve{u_1}{u_2}{u_3}$ and $\vv=\ve{v_1}{v_2}{v_3}$ we denote by $\uu \dd \vv$ the inner product, by $|\uu|$ the norm and by $\uu \times \vv$ the cross product. That is $\uu\dd \vv=u_1v_1+u_2v_2+u_3v_3$, $|\uu|=\sqrt{\uu\dd \uu}$ and $\uu \times \vv = (u_2v_3-u_3v_2)\ff_1+(u_3v_1-u_1v_3)\ff_2+(u_1v_2-u_2v_1)\ff_3$. We also recall the identities
 \begin{equation} \label{EqCross}
        (\mm \times \uu)\dd (\nn \times \vv)=(\mm\dd  \nn)(\uu \dd \vv)-(\uu \dd  \nn)(\vv \dd  \mm)
\end{equation}
and
 \begin{equation} \label{Eqcof}
  A\uu\times A\vv= \cof A (\uu \times \vv),
 \end{equation}
where $A$ is a $3 \times 3$ matrix. In particular, the matrix of cofactors, $\cof{A}\!$, measures how a vector normal to $\uu$ and $\vv$ deforms whenever $\uu$ and $\vv$ are deformed by $A$. If $A$ is invertible it holds that $\cof A=A^{-T}\det A $, where as usual $A^{-T}$ denotes the inverse of the transpose.

We end this section by summarising some important properties of rotation matrices, i.e. $3\times 3$ matrices $R$ such that $R^TR=\id$ and $\det R=1$. Any rotation matrix $R$ can be uniquely identified as a counterclockwise rotation by an angle $\phi$ about a vector $\uu$ and we write $R=R[\phi,\uu]$, where $\uu$ is always expressed in the standard basis $\ee_1=(1,0,0)^T$, $\ee_2=(0,1,0)^T$, $\ee_3=(0,0,1)^T$. The magnitude of the angle of rotation is given by $|\phi|=\arccos ((\tr R-1)/2)$, where $\tr R = \sum_{i=1}^{3} R_{ii}$ is the trace of the matrix $R$ and the sign of $\phi$ is given by $\sgn (\phi)=\sgn((\nn \times R\,\nn)\dd\uu)$, where $\nn$ is any vector that is not parallel to the axis of rotation $\uu$. In particular, reversing the sign of the axis $\uu \rightarrow -\uu$ is equivalent to reversing the sign of the angle of rotation $\phi \rightarrow -\phi$. Finally, by $\PP$ we denote the group of rotations that map a cube to itself (see \hyperref[App]{Appendix}) and we call two vectors $\nn, \nn'
$ \emph{crystallographically equivalent} iff $\nn'=P\nn$ for some $P\in \PP$.

\section[A unified approach]{A unified approach to phase transformation models in steels}\label{SecUnif}
Since Bain's seminal paper \cite{Bain} (see also \cite{OptLat} for a rigorous mathematical justification) it is well known that the pure stretches required to transform an fcc lattice to a bcc/bct lattice are given by the three Bain strains
 \begin{equation}\label{EqBain}
  B_1=\begin{pmatrix} \beta & 0 & 0 \\ 0 & {\al}& 0\\ 0 & 0 & {\al} \end{pmatrix},\, B_2=\begin{pmatrix} {\al} & 0 & 0 \\ 0 & \be & 0\\ 0 & 0 & {\al} \end{pmatrix},\, B_3=\begin{pmatrix} {\al} & 0 & 0 \\ 0 & {\al} & 0\\ 0 & 0 & \be \end{pmatrix},
 \end{equation}
where ${\al}=\frac{\sqrt{2}a}{a_0}$ and $\be=\frac{c}{a_0}$. Here $a_0$ is the lattice parameter of the fcc phase and $c \geq a$ are the lattice parameters of the bct phase ($a=c$ for bcc). An additional rigid body rotation $R$ does not change the bcc/bct lattice structure and hence any lattice transformation $T$ from fcc to bcc/bct is of the form
\begin{equation*}
 T=RB_i \mbox{ for some }i=1,2,3.
\end{equation*}
Now suppose that the transformation $T$ leaves a plane with normal $\nn$ and a direction $\vv$ within that plane unrotated, i.e. 
\begin{equation}\label{EqRel}
 \frac{\cof T \nn}{|\cof T\nn|}=R\frac{\cof{B_i} \nn}{|\cof{B_i}\nn|}=\nn \aand \frac{ T \vv}{| T\vv|}=R \frac{ B_i \vv}{| B_i\vv|}=\vv.
\end{equation}
Defining $\mm_i={\cof{B_i} \nn}/{|\cof{B_i}\nn|}$, $\uu_i={B_i \vv}/{| B_i\vv|}$, we observe that
\begin{equation*}
 \mm_i\dd \uu_i \propto \cof{B_i} \nn \dd B_i \vv=B_i^T\cof{B_i} \nn \dd  \vv\propto \nn \dd  \vv=0,\footnote{Recall that $x\propto y$ if there is a constant $c$ such that $x=cy$.}
\end{equation*}
where we have used that $\cof{B_i} \propto B_i^{-T}$ and that $\vv \perp \nn$. In particular, the pairs $\mm_i, \uu_i$ and $\nn, \vv$ are both orthonormal and thus there is a unique rotation $R=R_i$ such that $R_i\mm_i=\nn$ and $R_i\uu_i=\vv$ given by
\begin{align}\label{EqR}
 R_i=\begin{pmatrix}
 \vline & \vline & \vline \\
\nn & \vv & \nn\times \vv \\
\vline & \vline & \vline
\end{pmatrix}\begin{pmatrix}
 \hl \hfill \mm_i \hfill \hl \\
\hl \hfill \uu_i \hfill \hl \\
\hl \hfill  \mm_i \times \uu_i  \hfill \hl
\end{pmatrix}.
\end{align}
Consequently, for each $i=1,2,3$ there is exactly one transformation strain, $T_i=R_iB_i$, from fcc to bcc/bct that leaves the plane with normal $\nn$ and the direction $\vv$ within that plane unrotated. 

\subsubsection*{Identifying strains with orientation relationships}
\begin{figure}[h]
\captionsetup{format =plain}%
  \centering
  \includegraphics[width=9cm]{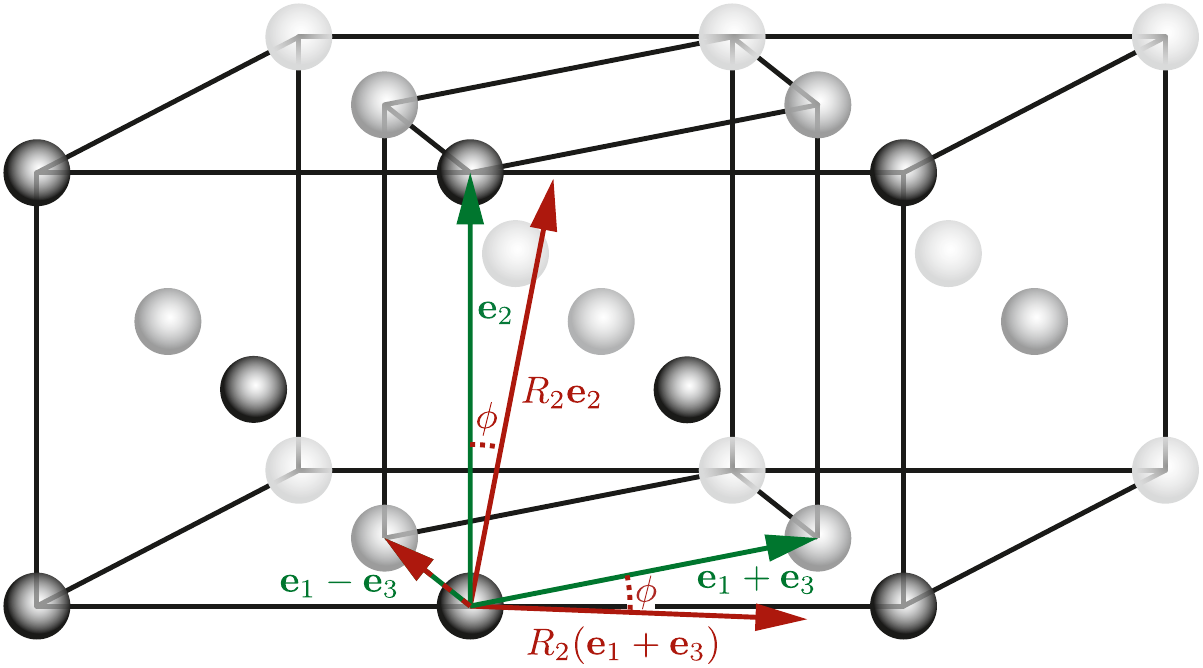}
  \caption{The green vectors $\ee_1-\ee_3,\ee_2,\ee_1+\ee_3$ are along the edges of the tetragonal bct cell that is contained in the fcc lattice and the red vectors are obtained through the rotation $R_2$.}\label{FigRotBain}
\end{figure}

Given the transformation strain $T_i$, we show how to compute the corresponding orientation relationship (OR). For simplicity, we focus on the case $i=2$; the remaining two cases can be treated analogously. From the pure Bain mechanism it is clear that the transformation $B_2$ results in a bcc/bct unit cell with edges along the directions $\ee_1-\ee_3$, $\ee_2$ and $\ee_1+\ee_3$ (see Figure~\ref{FigRotBain}). The additional rotation $R_2$ in the transformation $T_2$ then results in a bcc/bct unit cell with edges along the directions
\begin{equation*}
 R_2(\ee_1-\ee_3), R_2\ee_2 \aand R_2(\ee_1+\ee_3),
\end{equation*}
which form the natural basis for the bcc/bct lattice.
Noting that $\ee_1-\ee_3=R[45^\circ,\ee_2]\ee_1$ and $\ee_1+\ee_3=R[45^\circ,\ee_2]\ee_3$ we see that the change of basis matrix between fcc and bcc/bct is given by
$
 R_2 R[45^\circ,\ee_2],
$
i.e. $\mathbf{x}=\ve{x_1}{x_2}{x_3}_\gamma=\ve{\hat x_1}{\hat x_2}{\hat x_3}_{\al'}$, where   
\begin{equation}\label{EqTrafo0}
\begin{pmatrix}\hat x_1\\\hat x_2\\\hat x_3\end{pmatrix}=R[-45^\circ,\ee_2]R_2^T\begin{pmatrix}x_1\\x_2\\x_3\end{pmatrix}=:O_2 \begin{pmatrix}x_1\\x_2\\x_3\end{pmatrix}.
\end{equation}
In particular, through the matrix $O_2=R[-45^\circ,\ee_2]R_2^T$ one can express the coordinates of the unrotated plane $\nn$ and direction $\vv$ in the new bcc/bct (${\al'}$-) basis and hence determine the \emph{orientation relationship}. In general, the orientation relationship corresponding to $T_i=R_iB_i$ is given through the matrix
\begin{equation}\label{EqGenOr}
 O_i=R[-45^\circ,\ee_i]R_i^T,
\end{equation}
which we henceforth call the \emph{orientation relationship matrix}. We note that $R[45^\circ,\ee_i]=R[90^\circ,\ee_i]R[-45^\circ,\ee_i]$ with $R[90^\circ,\ee_i]\in \PP$, i.e. choosing the opposite sign for the $45^\circ$ rotation about $\ee_i$ simply leads to a crystallographically equivalent normal and direction. In summary, starting from the transformation $T_i$, we obtain the \emph{orientation relationship}
\begin{equation} \label{EqOR}
 \no{n_1}{n_2}{n_3}_\gamma \parallel \no{\hat n_1}{\hat n_2}{\hat n_3}_{\al'} \aand \ve{v_1}{v_2}{v_3}_\gamma \parallel \ve{\hat v_1}{\hat v_2}{\hat v_3}_{\al'},
\end{equation}
where the coordinates $\hat n_i$ and $\hat v_i$ are obtained by using the \emph{orientation relationship matrix} $O_i$ from \eqref{EqGenOr} in \eqref{EqTrafo0}.

Conversely, suppose that an OR of the form \eqref{EqOR} is given with the property that the normal $\no{n_1}{n_2}{n_3}_\gamma$ and the direction $\ve{v_1}{v_2}{v_3}_\gamma$ are left unrotated by the transformation. By the above process, we can compute three possible transformation strains $T_i$ and corresponding OR matrices $O_i$. For each OR matrix $O_i$ we can calculate the bcc/bct coordinates of $\no{n_1}{n_2}{n_3}_\gamma$ and $\ve{v_1}{v_2}{v_3}_\gamma$. For one of the matrices $O_i$, the calculated coordinates must agree, up to crystallographic equivalence, with the given OR and, hence, we may uniquely identify the Bain variant $B_i$, and the corresponding transformation strain $T_i$, that gives rise to the OR. If the coordinates do not agree for any $O_i$, then the OR cannot be compatible with the Bain mechanism.

\subsubsection*{Generating variants through crystallographic equivalence in the $\gamma$ phase}
Given a transformation strain $T$ (or equivalently the corresponding OR matrix $O$) we are able to generate further variants of $T$ through the application of $\PP$ in the reference configuration. To this end, we recall that given the fcc basis $\lbrace \ee_1,\ee_2, \ee_3\rbrace$, all crystallographically equivalent fcc bases are given by $\lbrace P\ee_1,P\ee_2, P\ee_3\rbrace$ for $P \in \PP$. Thus, letting $T$ as in \eqref{EqRel} and using the identity $P_i^TP_i=\id$ we infer that
\begin{equation*}
 \frac{\cof{(P_iTP_i^T)} P_i\nn}{|\cof {T\nn|}}=P_i\nn \aand \frac{  (P_iT P_i^T) P_i\vv}{| T\vv|}=P_i\vv.
\end{equation*}
That is, for each $i=1,2, \dotsc, 24$, the deformation $P_iTP_i^T$ leaves the plane with normal $P_i\nn$ and the direction $P_i\vv$ within that plane unrotated and thus describes a \emph{strain variant} of the original transformation strain $T$. Similarly, $P_iOP_i^T$ describes the corresponding \emph{orientation relationship variant}. We note that in general, it may happen (see e.g. the \NW\ model) that $P_iTP_i^T=P_jTP_j^T$ for some $i \neq j$ and thus there can be less than 24 distinct  variants for a given transformation strain (or equivalently for a given OR).

\section{The NW and KS models} \label{SecDer}
In this section, we derive the \NW\ and \KS\ models. Both models have the attractive feature of leaving a close-packed $\{1\,1\,1\}_\gamma$ plane and a close-packed $\langle \bar{1}\,1\,0\rangle$ direction within that plane unrotated. Owing to this feature they seem to be the most natural candidates for OR models. 
To carry out the derivation we apply our unified approach from Section~\ref{SecUnif} with
\[
\nn=\no{1}{1}{1}_\gamma \aand \vv=\ve{1}{0}{\bar1}_\gamma.
\]
\subsection*{The transformation with stretch component $\mathbf B_2$}
Let us consider the second Bain variant $B_2$. Noting that $\vv$ is an eigenvector of $B_2$, we immediately deduce that, by \eqref{EqRel}, $R_2\vv=\vv$ and thus $\vv$ is the axis of rotation. Regarding the  
angle of rotation we calculate
\begin{equation*}
 \tr R_2=\mm_2\dd \nn + \uu_2 \dd \vv + (\mm_2 \times \uu_2) \dd (\nn \times \vv)=2\mm_2\dd \nn+1,
\end{equation*}
where we used that $\uu_2=\vv$ and \eqref{EqCross}. Hence, the angle of rotation is given by
\begin{alignat}{2} 
  \arccos\left(\frac{\cof{B_2} \nn \dd \nn}{|\cof{B_2} \nn|}\right)\sgn((\mm_2 \times \nn)\dd v)
  &=\arccos \left(\frac{1+\sqrt{2}r}{ \sqrt{3}\sqrt{1+r^2}}\right)=:\phi(r), \label{EqPhi}
\end{alignat}
where $r=c/a=\sqrt{2}\be/{\al}$ is the ratio of tetragonality of the bct cell. In particular, for $r=1$ corresponding to a bcc product lattice we obtain $\phi(1)=\arccos\left(\frac{1+\sqrt{2}}{\sqrt{6}}\right)\approx 9.7356^\circ$.

Hence, the only transformation from fcc to bcc/bct with stretch component $B_2$ which leaves the plane $\no{1}{1}{1}_\gamma$ and the direction $\ve{ 1}{ 0}{\bar 1}_\gamma$ unrotated is 
\begin{equation}\label{EqT3}
 T_2=R_2\,B_2=R[\phi(r),\ve{ 1}{ 0}{\bar 1}]\, B_2.
\end{equation}
Regarding the orientation relationships corresponding to $T_2$, through \eqref{EqT3} and \eqref{EqGenOr}, we infer that 
$O_2 = R[-45^\circ,\ee_2]\,R[-\phi(r),\ve{ 1}{ 0}{\bar 1}]$ (cf. Figure~\ref{FigRotBain}). Consequently,
\begin{equation}\label{EqNW1nv}
\no{1}{1}{1}_\gamma \parallel \no{0}{1}{r}_{\al'} \aand \ve{1}{0}{\bar 1}_\gamma \parallel \ve{ 1}{0}{0}_{\al'}.
\end{equation}
Note that, as expected, the latter is a closest packed plane in the resulting bct lattice containing the bct direction $\ve{ 1}{0}{0}_{\al'}$.
Thus for $r=1$ (bcc) the transformation $T_2$ gives rise to the OR $\NW1$ (see Table~\ref{TableNWOR}) and henceforth we denote $T_2=T_{\NW1}$. The OR matrix $O_{\NW1}$ between fcc and bcc is given by
\begin{equation*}
 O_{\NW1}=R[-45^\circ,\ee_2]\,R[-9.7356^\circ,\ve{ 1}{ 0}{\bar 1}]\approx \left(
\begin{array}{ccc}
 0.7071 & 0 & -0.7071 \\
 0.1196 & 0.9856 & 0.1196 \\
 0.6969 & -0.1691 & 0.6969
\end{array}
\right),
\end{equation*}
and the corresponding transformation $T_{\NW1}$ is given by
\begin{equation*}
 T_{\NW1}=R[9.7356^\circ,\ve{ 1}{ 0}{\bar 1}]\, B_2\approx \left(
\begin{array}{ccc}
 1.1144 & 0.0949 & -0.0081  \\
  -0.1342 & 0.7823 & -0.1342 \\
 -0.0081  & 0.0949 & 1.1144
\end{array}
\right).
\end{equation*}
%

Next, we characterize the remaining \NW\ variants. Following our unified approach, they are given by $P_iT_{\NW 1}P_i^T$. Since $T_{\NW 1}=R[\phi(r),\ve{ 1}{ 0}{\bar 1}]\, B_2$, $P_2\ve{ 1}{ 0}{\bar 1}_\gamma=\ve{ 1}{ 0}{\bar 1}_\gamma$ and $P_2B_2P_2^T=B_2$ we deduce that $P_2T_{\NW 1}P_2^T=T_{\NW 1}$ and similarly that $P_{2j}T_{\NW 1}P_{2j}^T=P_{2j-1}T_{\NW 1}P_{2j-1}^T $ for any $j=2,\dotsc,12$. Thus there are only 12 \NW\ strain variants given by 
\begin{equation*}
T_{\NW j}:= P_{2j-1}T_{\NW 1}P_{2j-1}^T = R[\phi(r),P_{2j-1}\ve{ 1}{ 0}{\bar 1}]\,P_{2j-1}B_2P_{2j-1}^T,
\end{equation*}
for $j=1,2,\dotsc, 12$. In particular, $T_{\NW j}$ has a stretch component $P_{2j-1}B_2P_{2j-1}^T$ followed by a rotation of $\phi(r)$ about $P_{2j-1} \ve{ 1}{ 0}{\bar 1}_\gamma$. The corresponding OR matrices are obtained by the same conjugation. That is
\begin{equation*}
 O_{\NW j}=P_{2j-1}O_{\NW 1}P_{2j-1}^T=R[-45^\circ,P_{2j-1}\ee_2]\,R[-\phi(r),P_{2j-1}\ve{ 1}{ 0}{\bar 1}],
\end{equation*}
for $j=1,2,\dotsc, 12$. Thus, by \eqref{EqNW1nv}, $O_{\NW j}$ maps the fcc normal $P_{2j-1}\nn$ and fcc vector $P_{2j-1}\vv$ to the bcc/bct normal $P_{2j-1}\no{0}{1}{r}_{\al'}$ and the bcc/bct direction $P_{2j-1}\ve{1}{0}{0}_{\al'}$ 
(see Table~\ref{TableNWORA} in the Appendix). It is easy to verify that, for $r=1$, the resulting bcc vectors are crystallographically equivalent (through $P_{2j-1}^T$) to the bcc vector $\ve{1}{0}{0}_{\al'}$ and the bcc normal $\no{0}{1}{1}_{\al'}$, giving the \NW\ variants as in Table~\ref{TableNWOR}. We note that the choice of sign for the 45$^\circ$ rotation about $\ee_2$, as well as the enumeration of $\PP$, has been carefully made so that the OR $\NW j$ is obtained through $P_{2j-1}^T$. A choice of the opposite sign and/or a different enumeration of $\PP$, will not alter the result but will lead to bcc/bct coordinates that are crystallographically equivalent to the ones in Table~\ref{TableNWOR} through 
different 
elements of $\PP$. 

 \begin{center}
 {\singlespacing
\begin{threeparttable}[ht] \captionsetup{format =plain}%
\caption{The NW orientation relationships. The corresponding variants in each row are given by 
$T_{\NW j}=R[\phi(r),\vv_j]\, B_j$.}\label{TableNWOR}

 \begin{tabular}{cccccc}\hline
  O.R.\tnote{a} &  fcc plane\tnote{b} & bcc plane & fcc direction\tnote{c} & bcc direction& Bain Variant\tnote{d} \\
\hline    & & & & \vspace*{-1em} \\
 \NW1 &  $\no{1}{1}{1}_{\gamma}$ & $\no{0}{1}{1}_{\al'}$ & $\ve{1}{0}{\bar 1}_{\gamma}$ & $\ve{1}{0}{0}_{\al'}$  & $B_2$\\ 
 \NW2 &  $\no{1}{1}{1}_{\gamma}$ & $\no{0}{1}{1}_{\al'}$ & $\ve{ \bar 1}{1}{0}_{\gamma}$ & $\ve{1}{0}{0}_{\al'}$ &$B_3$ \\
 \NW3 &  $\no{1}{1}{1}_{\gamma}$ & $\no{0}{1}{1}_{\al'}$ & $\ve{0}{\bar1}{1}_{\gamma}$ & $\ve{1}{0}{0}_{\al'}$  &$B_1$\\
  & & & & \\
    \NW4 & $\no{\bar{1}}{1}{1}_{\gamma}$ & $\no{0}{1}{1}_{\al'}$ &  $\ve{1}{0}{1}_{\gamma}$& $\ve{1}{0}{0}_{\al'}$ & $B_2$\\
    \NW5 &  $\no{\bar{1}}{1}{1}_{\gamma}$ & $\no{0}{1}{1}_{\al'}$ & $\ve{\bar1}{\bar 1}{0}_{\gamma}$ & $\ve{1}{0}{0}_{\al'}$ &$B_3$\\
  \NW6 &  $\no{\bar{1}}{1}{1}_{\gamma}$ & $\no{0}{1}{1}_{\al'}$ &  $\ve{0}{1}{\bar1}_{\gamma}$& $\ve{1}{0}{0}_{\al'}$&$B_1$ \\
     & & & & \\ 
     \NW7 &  $\no{1}{\bar{1}}{1}_{\gamma}$ & $\no{0}{1}{1}_{\al'}$ & $\ve{\bar1}{0}{1}_{\gamma}$ &$ \ve{1}{0}{0}_{\al'}$& $B_2$\\
  \NW8 & $\no{1}{\bar{1}}{{1}}_{\gamma}$ & $\no{0}{1}{1}_{\al'}$ &$ \ve{1}{1}{0}_{\gamma}$&$ \ve{1}{0}{0}_{\al'}$&$B_3$ \\
   \NW9 &  $\no{1}{\bar{1}}{1}_{\gamma}$ & $\no{0}{1}{1}_{\al'}$ &$ \ve{0}{\bar1}{\bar 1}_{\gamma}$ & $\ve{1}{0}{0}_{\al'}$ &$B_1$ \\
    & & & & \\     
 \NW10 &  $\no{1}{1}{\bar{1}}_{\gamma}$ & $\no{0}{1}{1}_{\al'}$ & $\ve{\bar 1}{0}{\bar 1}_{\gamma}$ &$ \ve{1}{0}{0}_{\al'}$ & $B_2$\\
  \NW11 &  $\no{1}{1}{\bar{1}}_{\gamma}$ & $\no{0}{1}{1}_{\al'}$ & $\ve{1}{\bar 1}{0}_{\gamma}$&$\ve{1}{0}{0}_{\al'}$ &$B_3$\\
  \NW12 & $\no{1}{1}{\bar{1}}_{\gamma}$ & $\no{0}{1}{1}_{\al'}$ & $ \ve{0}{1}{1}_{\gamma}$&$\ve{1}{0}{0}_{\al'}$&$B_1$ \\
\hline
\end{tabular}
\begin{tablenotes}
\item[a]\hypertarget{fa}{$\NW j$}
\item[b] \hypertarget{fb}{$P_{2j-1}\no{1}{1}{1}_\gamma$}
\item[c] \hypertarget{fc}{$\vv_j=P_{2j-1}\ve{1}{0}{\bar 1}_\gamma$}
\item[d] \hypertarget{fd}{$B_j=P_{2j-1}B_2P_{2j-1}^T$}
\end{tablenotes}
\end{threeparttable}}
\end{center}
\subsection*{The transformation with stretch component $\mathbf B_3$}
 Similarly, using $B_3$ instead of $B_2$ in \eqref{EqRel} gives rise to a rotation $R_3$ satisfying
\begin{equation}\label{EqKS1}
 R_3\mm_3=\nn \aand R_3\uu_3=\vv.
\end{equation}
Noting that $R_{\NW2}\mm_3=\nn$ we immediately see that $R_3\,R_{\NW2}^T\nn=\nn$ and
\begin{equation*}
 R_3=R[\theta,\nn]\,R_{\NW2}=R[\theta,\ve{1}{1}{1}]\,R[\phi(r),\ve{\bar 1}{ 1}{0}]
\end{equation*}
for some angle $\theta=\theta(r)$. Let us first determine the sign of $\theta(r)$. By 
\eqref{EqKS1}, we have that $R[\theta, \nn]\,R_{\NW2}\uu_3=\vv$ and thus
$
 \sgn \theta(r)=\sgn (R_{\NW2}\uu_3\times \vv)\cdot \nn=1.
 $
For the angle itself we deduce from \eqref{EqR} that
\begin{equation}\label{EqTheta}
 \theta(r)=\arccos \left(\frac{\tr R[\theta, \nn]-1}{2}\right)= \arccos \left(\frac{\sqrt{3} \sqrt{r^2+1}+1}{2 \sqrt{r^2+2}}\right).
\end{equation}
For $r=1$ (bcc) this angle is given by $\theta(1)=\arccos\left(\frac{1+\sqrt{6}}{2\sqrt{3}}\right)\approx5.2644^\circ$. Hence, the only transformation from fcc to bcc/bct with stretch component $B_3$ which leaves the plane $\no{1}{1}{1}_\gamma$ and the direction $\ve{ 1}{ 0}{\bar 1}_\gamma$ unrotated is 
\begin{equation}\label{EqTKS1}
 T_3=R_3\,B_3=R[\theta(r), \ve{1}{1}{1}]\,R[\phi(r),\ve{\bar 1 }{ 1}{0}]\, B_3.
\end{equation}
Regarding the corresponding orientation relationships, by \eqref{EqGenOr}, we deduce that 
\begin{equation}\label{EqKS1n}
 O_3 = R[45^\circ,\ee_3]\,R[-\phi(r),\ve{\bar 1 }{ 1}{0}]\,R[-\theta(r), \ve{1}{1}{1}]
\end{equation}
and, consequently,
\begin{equation} \label{EqKS1nv}
\no{1}{1}{1}_\gamma \parallel \no{0}{r}{1}_{\al'} \aand \ve{1}{0}{\bar 1}_\gamma \parallel \ve{ 1}{1}{\bar r}_{\al'}.
\end{equation}
These correspond to a closest packed plane in the resulting bcc/bct lattice and the close packed direction in that plane. Clearly, for $r=1$ (bcc), the transformation $T_3$ gives rise to the OR $\KS1$ (see Table~\ref{TableKSOR}) and henceforth we denote $T_3=T_{\KS1}$. The OR matrix $O_{\KS1}$ between fcc and bcc is then given by
\begin{align*}
O_{\KS1}&=R[45^\circ,\ee_3]\,R[-9.7356^\circ,\ve{\bar 1 }{ 1}{0}]\,R[-5.2644^\circ, \ve{1}{1}{1}]\\
&\approx \left(
\begin{array}{ccc}
0.7416 & -0.6667 & -0.0749 \\
0.6498 &  0.7416 &-0.1667  \\
0.1667 & 0.07492 & 0.9832 
\end{array}
\right)
\end{align*}
and the transformation strain by
\begin{equation*}
 T_{\KS1}=R[5.2644^\circ, \ve{1}{1}{1}]\,R[9.7356^\circ,\ve{\bar 1 }{ 1}{0}]\, B_3\approx \left(
\begin{array}{ccc}
  1.1044 & -0.0728 & 0.1323 \\
 0.0595 & 1.1177 &  0.0595\\
  -0.1917 & -0.0728 & 0.7803
\end{array}
\right).
\end{equation*}
The remaining \KS\ strain variants are $T_{\KS j}:=P_jT_{\KS1}P_j^T$ and by \eqref{EqTKS1} they are given by
\begin{equation*}
 T_{\KS j}=R[\theta(r), P_j\ve{1}{1}{1}]\,R[\phi(r),P_j\ve{\bar 1 }{ 1}{0}]\,P_jB_3P_j^T.
\end{equation*}
In particular, $T_{\KS j}$ leaves the close packed plane $P_j\nn$ and the close packed direction $P_j\vv$ within that plane unrotated. The corresponding OR variants are given by $O_{\KS j}=P_{j}O_{\KS 1}P_{j}^T$ and $O_{\KS j}$ maps the fcc normal $P_{j}\nn$ and fcc direction $P_{j}\vv$ to the bcc/bct normal $P_{j}\no{0}{r}{1}_{\al'}$ and the bcc/bct direction $P_{j}\ve{1}{1}{\bar r}_{\al'}$ (see Table~\ref{TableKSORA} in the Appendix).
\subsection*{The transformation with stretch component $B_1$}
Let us, for example, consider $P=P_2$. Then
\begin{equation*}
 P_2\nn=-\nn, \,P_2\vv=\vv \aand P_2B_3P_2^T=B_1
\end{equation*}
and thus $T_{\KS 2}=R[-\theta(r), \ve{1}{1}{1}]\,R[\phi(r),\ve{\bar 1 }{ 1}{0}]\,B_1$ is the only transformation with stretch component $B_1$ that leaves the close packed plane $\no{1}{1}{1}_\gamma$ and the close packed direction $\ve{ 1}{ 0}{\bar 1}_\gamma$ unrotated. It is therefore the third and last solution of \eqref{EqRel}.

Just like in the derivation of the $\NW$ variants, care has been taken so that all odd $\KS(2j-1)$ variants correspond immediately to the entries in Table~\ref{TableKSOR} and the crystallographic equivalence in the bcc/bct lattice is given by $P_{2j-1}^T$. However, unlike the $\NW$ variants, $T_{\KS2}=P_2 T_{\KS1} P_2^T\neq T_{\KS1}$ are distinct and thus the ORs are different. To illustrate this, let us take $O_{\KS 2}=P_{2}O_{\KS 1}P_{2}^T$ and investigate its action on the fcc plane with normal $\nn=\no{1}{1}{1}_\gamma$ and the fcc direction $\vv=\ve{ 1}{ 0}{\bar 1}_\gamma$. We have
\begin{alignat}{2}\label{ORKS2}
 &O_{\KS 2}\nn=P_{2}O_{\KS 1}(-\nn)=-P_2 \no{0}{r}{1}_{\al'}= \no{1}{ r}{0}_{\al'}\\ \nonumber
 \aand &O_{\KS 2}\vv=P_{2}O_{\KS 1} \vv= P_2\ve{1}{1}{\bar r}_{\al'}=\ve{ r}{\bar 1}{\bar 1}_{\al'},
\end{alignat}
which are the closest packed plane and close packed direction in that plane in the resulting bct lattice. If $r=1$ (bcc), noting that $P_3 \no{1}{ r}{0}_{\al'}= \no{0}{1}{r}_{\al'}$ and $P_3 \ve{ r}{\bar 1}{\bar 1}_{\al'}=\ve{\bar 1}{r}{\bar 1}_{\al'}$ we obtain, up to crystallographic equivalence in the bcc lattice (by $P_3$)\footnote{Nevertheless, $P_3$ is not a lattice invariant rotation for the resulting bct lattice.} the OR associated to $\KS 2$ (cf. Table~\ref{TableKSOR}). The ORs for the remaining even $\KS(2j)$ are obtained analogously and the required crystallographic equivalence transformation in the bcc lattice is given by $P_3 P_{2j}^T$. Figure~\ref{FigVarRel} shows the relations between all Bain, \NW\ and \KS\ variants.
\begin{figure}[h]
\captionsetup{format =plain}%
\xymatrix@C=0.2em@R=2pt{
   &&&     &&& \KS{2} & &&&     &&& \KS{4} &     &&&&&& \KS{1}\\
    &&&    \NW3\ar[urrr] \ar[rrr] &&& \KS{3} & &&& \NW{1}\ar[urrr] \ar[rrr] &&& \KS{5} & &&& \NW{2}\ar[urrr] \ar[rrr] &&&  \KS{6}\\
  &&&    &&&  \KS{7} &  &&&    &&&  \KS{9} & &&&    &&&  \KS{8} \\
    &&&     \NW6\ar[urrr] \ar[rrr] &&&  \KS{10} &  &&&     \NW{4}\ar[urrr] \ar[rrr] &&&  \KS{12} & &&&     \NW{5}\ar[urrr] \ar[rrr] &&&  \KS{11} \\
  B_1 \ar[uuurrr]\ar[urrr]\ar[rrr]\ar[ddrrr]  &&&   \NW{9} \ar[drrr] \ar[rrr] &&&  \KS{13} & B_2 \ar[uuurrr]\ar[urrr]\ar[rrr]\ar[ddrrr]  &&&   \NW{7} \ar[drrr] \ar[rrr] &&&  \KS{15} & B_3 \ar[uuurrr]\ar[urrr]\ar[rrr]\ar[ddrrr]  &&&   \NW{8} \ar[drrr] \ar[rrr] &&&  \KS{14}\\
   &&&   &&&    \KS{16} & &&&   &&&    \KS{18} & &&&   &&&    \KS{17} \\
    &&&     \NW{12}\ar[drrr] \ar[rrr] &&&  \KS{19} & &&&     \NW{10}\ar[drrr] \ar[rrr] &&&  \KS{21} & &&&     \NW{11}\ar[drrr] \ar[rrr] &&&  \KS{20} \\
   &&&   &&&    \KS{22} & &&&   &&&    \KS{24} &  &&&    &&& \KS{23}\\
  }
  \caption{An arrow from a Bain variant $B_k$ to an \NW\ variant $\NW j$ signifies that $T_{\NW j}=R[\phi(r),\vv_j]B_k$ (cf. Table~\ref{TableNWOR}). Respectively, an arrow from an \NW\ variant $\NW j$
to a \KS\ variant $\KS i$ signifies that $T_{\KS i}=R[(-1)^{i+1}\theta(r),\nn_i]T_{\NW j}$ (cf. Table~\ref{TableKSOR}).}\label{FigVarRel}
\end{figure}
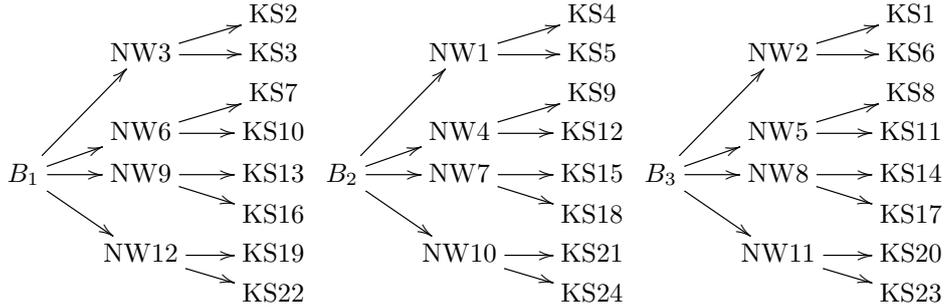
 \begin{center}
 {\singlespacing
\begin{threeparttable}[ht] \captionsetup{format =plain}
\caption{The \KS\ orientation relationships. The corresponding variants in each row are given by 
$T_{\KS j}=R[(-1)^{j+1}\theta(r), \nn_j]\,R[\phi(r),P_j\ve{\bar 1 }{ 1}{0}]\, B_j$.}\label{TableKSOR}
 \begin{tabular}{cccccc}\hline
  O.R.\tnote{a} &  fcc plane\tnote{b} & bcc plane & fcc direction\tnote{c} & bcc direction& Bain Variant\tnote{d} \\
\hline    & & & & \vspace*{-1em} \\
 $\KS1 $&$ \no{1}{1}{1}_{\gamma} $&$\no{0}{1}{1}_{\al'} $&$\ve{1}{0}{\bar 1}_{\gamma} $&$\ve{1}{1}{\bar 1}_{\al'}  $&$B_3$\\ 
 $\KS2 $&$ \no{1}{1}{1}_{\gamma} $&$\no{0}{1}{1}_{\al'} $&$\ve{1}{0}{\bar 1}_{\gamma} $&$\ve{\bar 1}{1}{\bar 1}_{\al'}$ &$B_1$\\
 $\KS3 $&$ \no{1}{1}{1}_{\gamma} $&$\no{0}{1}{1}_{\al'} $&$\ve{ \bar 1}{1}{0}_{\gamma} $&$\ve{1}{1}{\bar 1}_{\al'} $&$B_1 $\\
 $\KS4 $&$ \no{1}{1}{1}_{\gamma} $&$\no{0}{1}{1}_{\al'} $&$\ve{ \bar 1}{1}{0}_{\gamma} $&$\ve{\bar 1}{1}{\bar 1}_{\al'} $&$B_2$\\ 
 $\KS5 $&$ \no{1}{1}{1}_{\gamma} $&$\no{0}{1}{1}_{\al'} $&$\ve{0}{\bar1}{1}_{\gamma} $&$\ve{1}{1}{\bar 1}_{\al'}  $&$B_2$\\
 $\KS6 $&$ \no{1}{1}{1}_{\gamma} $&$\no{0}{1}{1}_{\al'} $&$\ve{0}{\bar1}{1}_{\gamma} $&$\ve{\bar 1}{1}{\bar 1}_{\al'}$&$B_3 $\\
  &&&&\\
    $\KS7 $&$\no{\bar{1}}{1}{1}_{\gamma} $&$\no{0}{1}{1}_{\al'} $&$ \ve{1}{0}{1}_{\gamma}$&$\ve{1}{1}{\bar 1}_{\al'}$&$B_1$\\
  $\KS8 $&$\no{\bar{1}}{1}{1}_{\gamma} $&$\no{0}{1}{1}_{\al'} $&$ \ve{1}{0}{1}_{\gamma}$&$\ve{\bar 1}{1}{\bar 1}_{\al'}$&$B_3 $\\
    $\KS9 $&$ \no{\bar{1}}{1}{1}_{\gamma} $&$\no{0}{1}{1}_{\al'} $&$\ve{\bar1}{\bar 1}{0}_{\gamma}$&$\ve{1}{1}{\bar 1}_{\al'} $&$B_2$\\
   $\KS10 $&$ \no{\bar{1}}{1}{1}_{\gamma} $&$\no{0}{1}{1}_{\al'} $&$\ve{\bar1}{\bar 1}{0}_{\gamma}$&$\ve{\bar 1}{1}{\bar 1}_{\al'}$&$B_1$\\
  $\KS11 $&$ \no{\bar{1}}{1}{1}_{\gamma} $&$\no{0}{1}{1}_{\al'} $&$ \ve{0}{1}{\bar1}_{\gamma}$&$\ve{1}{1}{\bar 1}_{\al'}$&$B_3 $\\
  $\KS12 $&$ \no{\bar{1}}{1}{1}_{\gamma} $&$\no{0}{1}{1}_{\al'} $&$\ve{0}{1}{\bar1}_{\gamma}$&$\ve{\bar 1}{1}{\bar 1}_{\al'}$&$B_2$\\
     &&&&\\ 
     $\KS13 $&$ \no{1}{\bar{1}}{1}_{\gamma} $&$\no{0}{1}{1}_{\al'} $&$\ve{\bar1}{0}{1}_{\gamma}$&$\ve{1}{1}{\bar 1}_{\al'}$&$B_1$\\
  $\KS14 $&$ \no{1}{\bar{1}}{1}_{\gamma} $&$\no{0}{1}{1}_{\al'} $&$ \ve{\bar1}{0}{1}_{\gamma}$&$\ve{\bar 1}{1}{\bar 1}_{\al'}$&$B_3 $\\
  $\KS15 $&$\no{1}{\bar{1}}{{1}}_{\gamma} $&$\no{0}{1}{1}_{\al'} $&$\ve{1}{1}{0}_{\gamma}$&$\ve{1}{1}{\bar 1}_{\al'}$&$B_2 $\\
    $\KS16 $&$\no{1}{\bar{1}}{{1}}_{\gamma} $&$\no{0}{1}{1}_{\al'} $&$ \ve{1}{1}{0}_{\gamma}$&$\ve{\bar 1}{1}{\bar 1}_{\al'}$&$B_1$\\
   $\KS17 $&$ \no{1}{\bar{1}}{1}_{\gamma} $&$\no{0}{1}{1}_{\al'} $&$\ve{0}{\bar1}{\bar 1}_{\gamma}$&$\ve{1}{1}{\bar 1}_{\al'}$&$B_3 $\\
  $\KS18 $&$ \no{1}{\bar1}{1}_{\gamma} $&$\no{0}{1}{1}_{\al'} $&$\ve{0}{\bar1}{\bar 1}_{\gamma}$&$\ve{\bar 1}{1}{\bar 1}_{\al'}$&$B_2$\\
    &&&&\\     
 $\KS19 $&$ \no{1}{1}{\bar{1}}_{\gamma} $&$\no{0}{1}{1}_{\al'} $&$\ve{\bar 1}{0}{\bar 1}_{\gamma} $&$\ve{1}{1}{\bar 1}_{\al'} $&$B_1$\\
 $\KS20 $&$ \no{1}{1}{\bar{1}}_{\gamma} $&$\no{0}{1}{1}_{\al'} $&$\ve{\bar 1}{0}{\bar 1}_{\gamma}$&$\ve{\bar 1}{1}{\bar 1}_{\al'}$&$B_3 $\\
  $\KS21 $&$ \no{1}{1}{\bar{1}}_{\gamma} $&$\no{0}{1}{1}_{\al'} $&$\ve{1}{\bar 1}{0}_{\gamma}$&$\ve{1}{1}{\bar 1}_{\al'}$&$B_2$\\
  $\KS22 $&$ \no{1}{1}{\bar{1}}_{\gamma} $&$\no{0}{1}{1}_{\al'} $&$\ve{1}{\bar 1}{0}_{\gamma}$&$\ve{\bar 1}{1}{\bar 1}_{\al'}$&$B_1$\\
  $\KS23 $&$\no{1}{1}{\bar{1}}_{\gamma} $&$\no{0}{1}{1}_{\al'} $&$ \ve{0}{1}{1}_{\gamma}$&$\ve{1}{1}{\bar 1}_{\al'}$&$B_3 $\\
  $\KS24 $&$\no{1}{1}{\bar{1}}_{\gamma} $&$\no{0}{1}{1}_{\al'} $&$ \ve{0}{1}{1}_{\gamma}$&$\ve{\bar 1}{1}{\bar 1}_{\al'}$&$B_2 $\\
\hline
\end{tabular}
\begin{tablenotes}
\item[a]\hypertarget{ffa}{$\KS j$}
\item[b] \hypertarget{ffb}{$\nn_j=(-1)^{j+1}P_{j}\no{1}{1}{1}_{\gamma}$}
\item[c] \hypertarget{ffc}{$P_{j}\ve{1}{0}{\bar 1 }_{\gamma}$}
\item[d] \hypertarget{ffd}{$B_j=P_{j}B_3P_{j}^T$}
\end{tablenotes}
\end{threeparttable}}
\end{center}

\subsection{Relation to other descriptions}\label{SecOth}
 In the literature (see e.g. \cite{Kallend,Ray,Bunge}) the \NW\ ORs are sometimes described as $\zeta=\arccos\left(\tfrac{1}{\sqrt{6}}-\tfrac{1}{2}\right)\approx 95.264^\circ$ rotations about $\langle hkl\rangle$ where $\ve{h}{k}{l}=[1+\sqrt{2}+\sqrt{3},\sqrt{2},-1+\sqrt{2}+\sqrt{3}]$ and the \KS\ ORs as 90$^\circ$ rotations about $\langle 112\rangle$. We show that these descriptions follow, up to crystallographic equivalence, from the above derivation. 
Let us start with the OR for $\NW1$. With the choice $P_3=R[120^\circ,\ve{1}{1}{1}]$ we obtain
\begin{equation*}
 P_3O_{\NW1}=R[\zeta,\ve{h}{k}{l}]\approx R[95.264^\circ, ( 0.85, 0.29,0.44)]
\end{equation*}
and thus $P_{2j-1} P_3O_{\NW1} P_{2j-1}^T=P O_{\NW j}=R[\zeta,P_{2j-1}\ve{h}{k}{l}]$ for some\footnote{$P=P_3$ for $j\in \lbrace1,2,3\rbrace$, $P=P_{18}$ for $j\in \lbrace4,5,6\rbrace$, $P=P_{24}$ for $j\in \lbrace7,8,9\rbrace$ and $P=P_{12}$ for $j\in \lbrace10,11,12\rbrace$} $P\in \PP$. That is, up to crystallographic equivalence in the bcc lattice, $O_{\NW j}$ is a $\zeta\approx 95.264^\circ$ rotation about $P_{2j-1} \ve{h}{k}{l}$ (see Table~\ref{TableNWOther}).
\begin{table}[h]
\captionsetup{format =plain}%
\centering
\begin{tabular}{ccccc}
\cline{1-2} \cline{4-5}
 OR & OR matrix && OR & OR matrix  \\
\cline{1-2} \cline{4-5}
& & & & \vspace*{-1em} \\
\NW1 &  $R[95.264^\circ,\ve{h}{k}{l}]$ && \NW7 &  $R[95.264^\circ,\ve{l}{\bar k}{h}]$\\
\NW2 &  $R[95.264^\circ,\ve{l}{h}{k}]$ && \NW8 &  $R[95.264^\circ,\ve{h}{\bar l}{k}]$\\
\NW3 &  $R[95.264^\circ,\ve{k}{l}{h}]$ && \NW9 &  $R[95.264^\circ,\ve{k}{\bar h}{l}]$\\
  & & & \\
\NW4 &  $R[95.264^\circ,\ve{\bar l}{k}{h}]$ && \NW10 &  $R[95.264^\circ,\ve{l}{k}{\bar h}]$\\
\NW5 &  $R[95.264^\circ,\ve{\bar h}{l}{k}]$ && \NW11 &  $R[95.264^\circ,\ve{h}{l}{\bar k}]$\\
\NW6 &  $R[95.264^\circ,\ve{\bar k}{h}{l}]$ && \NW12 &  $R[95.264^\circ,\ve{k}{h}{\bar l}]$\\
\cline{1-2} \cline{4-5}
\end{tabular}
\caption{The OR matrices corresponding to the \NW\ orientation relationships. Here, $\ve{h}{k}{l}=[1+\sqrt{2}+\sqrt{3},\sqrt{2},-1+\sqrt{2}+\sqrt{3}]\approx (0.85, 0.29,0.44)$.}\label{TableNWOther}
\end{table}
Next, let us consider the OR for $\KS1$. With the choice $P_{10}=R[-120^\circ,\ve{1}{\bar 1}{1}]$ we obtain
\begin{equation*}
 P_{10}O_{\KS1}=R[90^\circ,\ve{\bar 1}{2}{\bar 1}]
\end{equation*}
and thus $P_{j} P_{10}O_{\KS1} P_{j}^T=P O_{\KS j}=R[90^\circ,P_j\ve{\bar 1}{2}{\bar 1}]$ for some\footnote{$P=P_jP_{10}P_j^T$} $P\in \PP$, i.e. up to crystallographic equivalence in the bcc lattice, $O_{\KS j}$ is a $90^\circ$ rotation about $P_{j} \ve{\bar 1}{2}{\bar 1}$ (see Table~\ref{TableKSOther}).

\begin{table}[h]
\centering
\captionsetup{format =plain}%
\begin{tabular}{rlcrl}
\cline{1-2} \cline{4-5}
 OR & OR matrix && OR & OR matrix  \\
\cline{1-2} \cline{4-5}
& & & & \vspace*{-1em} \\
\KS1 &  $R[+90^\circ,\ve{\bar 1}{2}{\bar 1}]$ && \KS13 &  $R[+90^\circ,\ve{\bar 1}{\bar 2 }{\bar 1}]$\\
\KS2 &  $R[-90^\circ,\ve{\bar 1}{2}{\bar 1}]$ && \KS14 &  $R[-90^\circ,\ve{\bar 1}{\bar 2 }{\bar 1}]$\\
\KS3 &  $R[+90^\circ,\ve{\bar 1}{\bar 1}{2}]$ && \KS15 &  $R[+90^\circ,\ve{\bar 1}{1 }{2}]$\\
\KS4 &  $R[-90^\circ,\ve{\bar 1}{\bar 1}{2}]$ && \KS16 &  $R[-90^\circ,\ve{\bar 1}{1 }{2}]$\\
\KS5 &  $R[+90^\circ,\ve{2 }{\bar1}{\bar1}]$ && \KS17 &  $R[+90^\circ,\ve{2}{1}{\bar1 }]$\\
\KS6 &  $R[-90^\circ,\ve{2 }{\bar1}{\bar1}]$ && \KS18 &  $R[-90^\circ,\ve{2}{1}{\bar1 }]$\\
& & &\\
\KS7 &  $R[+90^\circ,\ve{1}{2}{\bar1}]$ && \KS19 &  $R[+90^\circ,\ve{\bar1}{2 }{1}]$\\
\KS8 &  $R[-90^\circ,\ve{1}{2}{\bar1}]$ && \KS20 &  $R[-90^\circ,\ve{\bar1}{2 }{1}]$\\
\KS9 &  $R[+90^\circ,\ve{1}{\bar 1}{2}]$ && \KS21 &  $R[+90^\circ,\ve{\bar1}{\bar1 }{\bar2}]$\\
\KS10 &  $R[-90^\circ,\ve{1}{\bar 1}{2}]$ && \KS22 &  $R[-90^\circ,\ve{\bar1}{\bar1 }{\bar2}]$\\
\KS11 &  $R[+90^\circ,\ve{ \bar 2}{\bar 1}{\bar 1}]$ && \KS23 &  $R[+90^\circ,\ve{2}{\bar1}{1 }]$\\
\KS12 &  $R[-90^\circ,\ve{ \bar 2}{\bar 1}{\bar 1}]$ && \KS24 &  $R[-90^\circ,\ve{2}{\bar1}{1 }]$\\
\cline{1-2} \cline{4-5}
\end{tabular}
\caption{The OR matrices corresponding to the \KS\ orientation relationships.}\label{TableKSOther}
\end{table}

\subsection{Twin relationships between KS variants} \label{SecTwin}
The knowledge of the transformation strains allows one to unambiguously identify pairs of \KS\ variants $\KS k$ and $\KS l$ 
that are twin related, i.e. variant pairs whose relative deformation is an invariant plane strain. That is 
\begin{equation*}\nonumber
 T_{\KS k}=T_{\KS l}(\id+\mathbf \bbb\otimes\mm),
\end{equation*}
where $\bbb\otimes\mm$ is the $3 \times 3$ matrix with components $(\bbb\otimes\mm)_{ij}=b_i m_j$. In particular, this implies that a fully coherent interface of normal $\mm$ can be formed between the two phases. We show that this can only happen between the pairs ${\KS(2j-1)}$ and ${\KS(2j)}$ and whenever this is the case the lattices on either side of the interface are related by a $180^\circ$ rotation about the common invariant fcc direction $P_{2j-1}\ve{1}{0}{\bar 1}=\vv_j$ (cf. Table~\ref{TableNWOR}). We start with ${\KS 1}$ and assume that 
\begin{equation}
 M_i:=T_{\KS i}-T_{\KS 1}=P_i T_{\KS 1} P_i^T-T_{\KS 1}=\bbb \tp \mm. \label{EqMj}
\end{equation}
Whenever $P_i$ does not leave $\vv_1$ invariant we have $(T_{\KS i}-T_{\KS 1})\vv_1 \neq 0$ and $(T_{\KS i}-T_{\KS 1})P_i\vv_1\neq 0$ and thus $\mm \parallel  \vv_1 \times P_i\vv_1$. Similarly, whenever $P_i$ does not leave $\nn_1=\no{1}{1}{1}_\gamma$ invariant, i.e. $i\geq 7$, we have\footnote{For an invertible matrix $A$, $v$ is an eigenvector of $\cof{A}$ iff it is an eigenvector of $A^T$.} $M_i^T \nn_1 \neq 0$ and $M_i^T \nn_i \neq 0$ and thus $\bbb \parallel  \nn_1 \times \nn_i$, where $\nn_i:=P_i\nn_1$. Hence for $i\geq 7$ it holds that
\begin{equation*}\nonumber
 M_i^T \nn_i\propto \mm \tp (\nn_1 \times \nn_i)\nn_i=\ssp{(\nn_1 \times \nn_i)}{\nn_i}\mm=0
\end{equation*}
and thus, since $\nn_i$ is an eigenvector of $T^T_{\KS i}$, it must also be an eigenvector of $T^T_{\KS 1}$. However, we know that this can only be the case for $i\leq 6$ (cf. Table~\ref{TableKSOR}), a contradiction. For the remaining cases, i.e. $2\leq i\leq 6$, we have 
\begin{equation*}\nonumber
 M_i P_i\vv_1 \propto \bbb \ssp{\vv_1 \times P_i\vv_1}{P_i\vv_1}=0
\end{equation*}
and thus since $P_i\vv_1$ is an eigenvector of $T_{\KS i}$ it must also be an eigenvector of $T_{\KS 1}$ which is again, unless $i=2$, a contradiction. Finally, 
\begin{equation*}\nonumber
 T_{\KS 2}-T_{\KS 1}=P_2 T_{\KS 1} P_2^T-T_{\KS 1}=\tfrac{2^{1/6}}{\sqrt{3}} \vv_1 \tp \ve{1}{0}{1},
\end{equation*}
where $P_2$ is a $180^\circ$ rotation about the common fcc direction $\vv_1$. Through conjugation with $P_{2j-1}$ we obtain that the relative deformations between $T_{\KS2j-1}$ and $T_{\KS2j}=P_{2j-1}T_{\KS2}P_{2j-1}^T$ are also invariant plane strains.

\subsection{The influence of tetragonality on the orientation relationships}\label{SecCarb}
 For many compositions of steel the $\al'$-phase is not cubic ($r=1$) but slightly tetragonal ($r>1$). For instance, the addition of carbon leads to a ratio of tetragonality approximately given by
\begin{equation}\label{EqTet}
 r=\frac{c}{a}=1+0.045 \, \mbox{wt}\, \%\, \mathrm C,
\end{equation}
for $\mathrm C$ in the range $0.4$--$2$ wt$\,\%\,\mathrm C$ (see \cite{Roberts,Winchell}).\footnote{Related experiments on $\mathrm{Fe}$-$7\%\, \mathrm{Al}$-$\mathrm C$ in \cite{Wata} showed that the tetragonality does not increase for carbon above $2\%$.} Similarly, the addition of nitrogen instead of carbon leads to a tetragonality ratio of
\begin{equation*}
 r=\frac{c}{a}=0.995+0.0383 \, \mbox{wt}\, \% \,\mathrm N,
\end{equation*}
for $\mathrm N$ in the range $0.6$--$2.9$ wt$\,\% \,\mathrm N$ (after Fig. $2.2$ in \cite{Nishiyamabook}). For small carbon content and certain $\mathrm{Fe}$-$\mathrm{Ni}$ alloys, such as the $\mathrm{Fe}$-$30\%\,\mathrm{Ni}$ alloy investigated in \cite{Nishiyama} and \cite{Wassermann}, the $\al'$-phase is likely to be cubic, however, alloying additional elements such as $\mathrm{Cr, Mn}$ or $\mathrm{Ti}$ leads again to a tetragonal $\al'$-phase. 

\begin{figure}[h]
  \centering
  \captionsetup{format =plain}%
  \includegraphics[width=12cm]{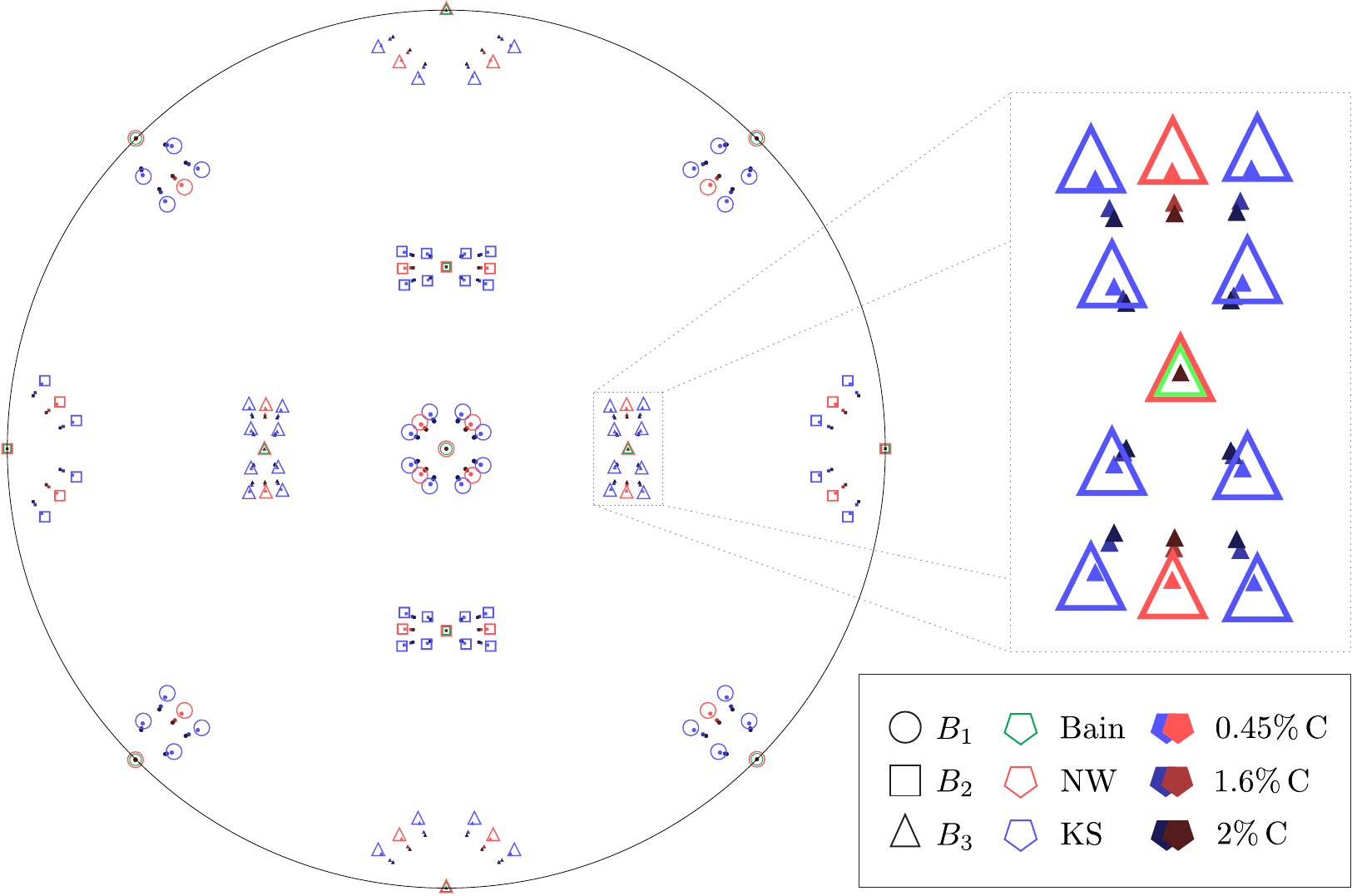}
  \caption{$\{1 0 0\}$ pole figures showing the change in the ORs with increasing carbon content. Hollow circles, squares and triangles correspond respectively to the fcc to bcc transformations with stretch components $B_1$, $B_2$ and $B_3$. The colours blue, red and green correspond respectively to \KS, \NW\ and Bain. The solid shapes correspond to increasing carbon content from lighter to darker shading and with values $0.45$, $1.6$ and $2$ wt $\,\%$ $\mathrm C$ respectively.}\label{FigCarbon}
\end{figure}

Our derivation in Section~\ref{SecDer} takes the tetragonality of the $\al'$-phase into account and the transformation strains, as well as the ORs, are derived for any ratio of tetragonality $1 \leq r < \sqrt{2}$.\footnote{Note that $r=\sqrt{2}$ corresponds to an fcc lattice and thus there is no phase transformation.} In particular, the angles of rotations $\phi(r)$ and $\theta(r)$ in \eqref{EqPhi} and \eqref{EqTheta} respectively decrease with increasing tetragonality and thus our theory predicts a narrower 
distribution of peaks in the pole figures. This prediction agrees very well with \cite{Ray} which summarises that \quo{investigators have shown that the chemical composition of steel has a significant effect on the nature and sharpness of the final transformation texture} and that increasing alloy content (i.e. higher tetragonality) leads to sharper textures (see e.g. \cite[Fig. 11-16]{Ray}).
Figure~\ref{FigCarbon} depicts the changes in the \NW\ and \KS\ ORs for different ratios of tetragonality obtained through \eqref{EqTet} for a carbon content increasing from $0\%$ to $2\%$. 
\section{Other orientation relationship models} \label{SecOtherORs}
In this section, we briefly comment on how our approach can be used to derive the Pitsch (see \cite{Pitsch}), Greninger-Troiano (\GT) (see \cite{GT}) and inverse Greninger-Troiano \GTp (see \cite{InverseGT}) OR models.

\subsection*{The Pitsch model}
Following \cite{Pitsch} the Pitsch ORs (\PT) are given as
\begin{equation}\label{Pitschnv}
 \no{1}{1}{0}_\gamma \parallel \no{\bar 1}{\bar 1}{\bar 2}\alp \aand \ve{0}{0}{1}_\gamma \parallel \ve{1}{\bar 1}{0}\alp.\footnote{In \cite{Pitsch} a third parallelism $\ve{1}{\bar 1}{0}_\gamma \parallel \ve{\bar 1}{\bar 1}{1}$ is provided, which is not required for our derivation but, nevertheless, follows from it.}
\end{equation}
Using our unified approach from Section~\ref{SecUnif} with $\nn=\no{1}{1}{0}$ and $\vv=\ve{1}{\bar 1}{0}$ we obtain $
 T_{\PT 1}=R[-\psi(r),\ve{0}{0}{1}]B_2$ and $O_{\PT 1}=R[-45^\circ,\ee_2]R[\psi(r),\ve{0}{0}{1}]$, where $\psi(r)=\arccos\left(\frac{\sqrt{2}+r}{\sqrt{2}\sqrt{2+r^2}}\right)$. The remaining eleven Pitsch OR and strain variants are given through conjugation with $\PP$. We note that for $r=1$, $\psi(1)=\phi(1)$, where $\phi(r)$ is given by \eqref{EqPhi} in the derivation of the \NW\ variants, and that $O_{\PT1}=O_{\NW 7}^T$ (similarly $O_{\PT j}=O_{\NW i}^T$ for some $i$). If instead of \eqref{Pitschnv} one uses the parallelisms $\no{0}{1}{0}_\gamma \parallel \no{1}{0}{1}\alp$ and $\ve{1}{0}{1}_\gamma \parallel \ve{\bar 1}{1}{1}\alp$ (as e.g. in \cite{InverseGT,Nolze}) the resulting strains and ORs are the same. Finally, we remark that occasionally \cite{PitschOdd} is also cited for the Pitsch ORs. However, the measurements in \cite{PitschOdd} are for cementite which has an orthorhombic crystal structure and thus our unified approach from Section~\ref{SecUnif} does not apply directly. Nevertheless, the 
underlying mechanism remains applicable if in \eqref{EqRel} one replaces the Bain strain by the respective strain required to transform austenite to cementite.
 
\subsection*{The Greninger-Troiano and inverse Greninger-Troiano models}
In \cite{GT}, Greninger-Troiano (\GT) studied a Fe-20\%Ni-0.8\%C crystal with $r=c/a=1.045$ and observed the following approximate parallelisms
\begin{equation}\nonumber
 \no{1}{1}{1}_\gamma :\no{1}{0}{1}\alp \approx 1^\circ, \, \veve{1}{1}{2}_\gamma: \ve{1}{0}{\bar 1}\alp\approx 2^\circ \aand \veve{1}{1}{0}_\gamma: \ve{1}{1}{\bar 1}\alp\approx 2.5^\circ.
\end{equation}
Apart from these original ORs (up to crystallographic equivalence), several authors use slightly different approximate parallelisms as defining features of the Greninger-Troiano (\GT) orientation relationships. For instance, \cite{Steels, Tsai} report $\nono{1}{1}{1}_\gamma:\nono{0}{1}{1}\alp \approx 0.2^\circ$ and $\veve{1}{0}{\bar 1}_\gamma : \veve{1}{1}{\bar 1}\alp\approx 2.7^\circ$ and \cite{InverseGT} uses the parallelisms
\begin{equation} \label{GT}
 \nono{1}{1}{1}_\gamma \parallel \nono{0}{1}{1}\alp \aand \veve{5}{12}{17}_\gamma \parallel \veve{7}{17}{17}\alp
\end{equation}
to approximate the \GT\ ORs. Using the parallelism condition \eqref{GT} our unified approach can capture the slight misorientations as an effect of the increased tetragonality of the bct lattice. With $\nn=\no{1}{1}{1}_\gamma$ and $\vv=\ve{\bar 5}{17}{\xo{12}}_\gamma$ we obtain $ T_{\GT 1}=R[\xi(r),\ve{1}{1}{1}]R[\phi(r),\ve{\bar 1}{1}{0}]B_3$ and $$O_{\GT1}=R[45^\circ,\ee_3]R[-\phi(r),\ve{\bar 1}{1}{0}]R[-\xi(r),\ve{1}{1}{1}]$$ with $\xi(r)=
\arccos\left(\frac{ 7^2+17^2 \sqrt{3}\sqrt{1 + r^2} }
 {\sqrt{2} \sqrt{5^2+12^2+17^2} \sqrt{7^2+17^2 + 17^2 r^2}}\right)$. In particular, we have $\no{1}{1}{1}_\gamma \parallel \no{0}{r}{1}_{\al'}$ and $\ve{\xo{12}}{\xo 5}{17}_\gamma \parallel \ve{\xo{7}}{\xo{17}}{17 r}_{\al'}$ and thus for the value $r=1.045$ studied in \cite{GT} we obtain $\no{1}{1}{1}_\gamma \parallel \no{0}{1.045}{1}_{\al'}:\no{0}{1}{1}\alp \approx 1.26^\circ$, $\ve{1}{1}{\bar{2}}_\gamma : \ve{0}{1}{\bar 1}\alp \approx 2.8^\circ$ and $\ve{1}{0}{\bar 1}_\gamma: \ve{1}{1}{\bar 1}\alp \approx 
2.9^\circ$.

The inverse \GT\ introduced in \cite{InverseGT} satisfy the conditions $ \no{\xo{17}}{\bar 7}{17}_\gamma \parallel \no{\bar 5}{\xo{12}}{17}\alp$ and $\ve{1}{0}{1}_\gamma \parallel \ve{1}{1}{1}\alp$ and as before our unified approach can be used to derive the corresponding strains and ORs. For further details on the \PT, \GT\, \GTp\ and also on the \NW\ and \KS\ ORs we refer to the Appendix.

\section{Conclusions}
A unified approach to derive transformation strains and orientation relationship models in steels is presented. An important aspect is the identification of strains with orientation relationships. The unified approach is used to derive the \NW, \KS\ and other models and extend them naturally to the situation of a tetragonal $\al'$ phase. The obtained dependence on the ratio of tetragonality seems to be in good qualitative agreement with experiments.\\

\section*{Acknowledgements} 
The research of A. M. leading to these results has received funding from the European Research Council under the European Union's $7^{\mbox{th}}$ Framework Programme (FP7/2007-2013) / ERC grant n$^\circ\, 291053$.

\newpage
\appendix
\setcounter{table}{0}
\renewcommand{\thetable}{A\arabic{table}}
\section{Overview of orientation relationship models} 
\addtocontents{toc}{\protect\setcounter{tocdepth}{1}}
\subsection{Nishiyama-Wassermann (NW)} 
\noindent The transformation $T_{\NW1}$ is uniquely defined through our unified approach (cf. Section~\ref{SecUnif}) as the transformation that:
\begin{itemize}
 \item leaves the normal $\nn=\no{1}{1}{1}_\gamma$ and the direction $\vv=\ve{1}{0}{\bar 1}_\gamma$ unrotated,
 \item has pure stretch component $B_2$.
\end{itemize}
The resulting transformation strain is
\begin{equation*}
 T_{\NW1}=R_2\,B_2=R[\phi(r),\ve{ 1}{ 0}{\bar 1}]\, B_2,
\end{equation*}
where $\phi(r)=\arccos \left(\frac{1+\sqrt{2}r}{ \sqrt{3}\sqrt{1+r^2}}\right)$. The corresponding OR matrix is 
$$O_{\NW1} = R[-45^\circ,\ee_2]\,R[-\phi(r),\ve{ 1}{ 0}{\bar 1}]$$
which yields the OR
\begin{equation*}
 \no{1}{1}{1}_\gamma \parallel \no{0}{1}{r}\alp \aand \ve{1}{0}{\bar 1}_\gamma \parallel \ve{ 1}{0}{0}\alp.
\end{equation*}
The application of $\PP$ yields the remaining eleven \NW\ ORs (cf. Table~\ref{TableNWORA}). Note that, unlike Table~\ref{TableNWOR}, Table~\ref{TableNWORA} takes the tetragonality of the bct lattice into account and the bct vectors are given in a way that is consistent with the transformation strains and not up to crystallographic equivalence.
 \begin{center}
 {\singlespacing
\begin{threeparttable}[ht] \captionsetup{format =plain}%
\caption{The NW orientation relationships. The corresponding transformation strain in each row is given by 
$T_{\NW j}=R[\phi(r),P_{2j-1}\ve{1}{0}{\bar 1}]\, B_j$.}\label{TableNWORA}

 \begin{tabular}{cccccc}\hline
  OR\tnote{a} &  fcc plane\tnote{b} & bcc plane\tnote{c} & fcc direction\tnote{d} & bcc direction\tnote{e}& Bain Variant\tnote{f} \\
\hline    & & & & \vspace*{-1em} \\
 \NW1 &  $\no{1}{1}{1}_\gamma $ & $\no{0}{1}{r}\alp$ & $\ve{1}{0}{\xo 1}_\gamma $ & $\ve{1}{0}{0}\alp$  & $B_2$\\ 
 \NW2 &  $\no{1}{1}{1}_\gamma $ & $\no{r}{0}{1}\alp$ & $\ve{ \xo 1}{1}{0}_\gamma $ & $\ve{0}{1}{0}\alp$ &$B_3$ \\
 \NW3 &  $\no{1}{1}{1}_\gamma $ & $\no{1}{r}{0}\alp$ & $\ve{0}{\xo 1}{1}_\gamma $ & $\ve{0}{0}{1}\alp$  &$B_1$\\
  & & & & \\
    \NW4 & $\no{\xo{1}}{1}{1}_\gamma $ & $\no{\xo r}{1}{0}\alp$ &  $\ve{1}{0}{1}_\gamma $& $\ve{0}{0}{1}\alp$ & $B_2$\\
    \NW5 &  $\no{\xo{1}}{1}{1}_\gamma $ & $\no{0}{r}{1}\alp$ & $\ve{\xo 1}{\xo 1}{0}_\gamma $ & $\ve{\xo 1}{0}{0}\alp$ &$B_3$\\
  \NW6 &  $\no{\xo{1}}{1}{1}_\gamma $ & $\no{\xo 1}{0}{r}\alp$ &  $\ve{0}{1}{\xo 1}_\gamma $& $\ve{0}{1}{0}\alp$&$B_1$ \\
     & & & & \\ 
     \NW7 &  $\no{1}{\xo{1}}{1}_\gamma $ & $\no{r}{\xo 1}{0}\alp$ & $\ve{\xo 1}{0}{1}_\gamma $ &$ \ve{0}{0}{1}\alp$& $B_2$\\
  \NW8 & $\no{1}{\xo{1}}{{1}}_\gamma $ & $\no{0}{\xo r }{1}\alp$ &$ \ve{1}{1}{0}_\gamma $&$ \ve{1}{0}{0}\alp$&$B_3$ \\
   \NW9 &  $\no{1}{\xo{1}}{1}_\gamma $ & $\no{1}{0}{\xo r}\alp$ &$ \ve{0}{\xo 1}{\xo 1}_\gamma $ & $\ve{0}{\xo 1}{0}\alp$ &$B_1$ \\
    & & & & \\     
 \NW10 &  $\no{1}{1}{\xo{1}}_\gamma $ & $\no{r}{1}{0}\alp$ & $\ve{\xo 1}{0}{\xo 1}_\gamma $ &$ \ve{0}{0}{\xo 1}\alp$ & $B_2$\\
  \NW11 &  $\no{1}{1}{\xo{1}}_\gamma $ & $\no{0}{r}{\xo 1}\alp$ & $\ve{1}{\xo 1}{0}_\gamma $&$\ve{1}{0}{0}\alp$ &$B_3$\\
  \NW12 & $\no{1}{1}{\xo{1}}_\gamma $ & $\no{1}{0}{\xo r}\alp$ & $ \ve{0}{1}{1}_\gamma $&$\ve{0}{1}{0}\alp$&$B_1$ \\
\hline
\end{tabular}
\begin{tablenotes}
\item[a] $\NW j$
\item[b] $P_{2j-1}\no{1}{1}{1}_\gamma$
\item[c] $P_{2j-1}\no{0}{1}{r}\alp$
\item[d] $P_{2j-1}\ve{1}{0}{\xo 1}_\gamma$
\item[e] $P_{2j-1}\no{1}{0}{0}\alp$
\item[f] $B_j=P_{2j-1}B_2P_{2j-1}^T$
\end{tablenotes}
\end{threeparttable}}
\end{center}

\subsection{Kurdjumov-Sachs (KS)}
\noindent The transformation $T_{\KS1}$ is uniquely defined through our unified approach (cf. Section~\ref{SecUnif}) as the transformation that:
\begin{itemize}
 \item leaves the normal $\nn=\no{1}{1}{1}_\gamma$ and the direction $\vv=\ve{1}{0}{\bar 1}_\gamma$ unrotated,
 \item has pure stretch component $B_3$.
\end{itemize}
The resulting transformation strain is
\begin{equation*}
 T_{\KS1}=R[\theta(r), \ve{1}{1}{1}]\,R[\phi(r),\ve{\bar 1 }{ 1}{0}]\, B_3,
\end{equation*}
where $\theta(r)= \arccos \left(\frac{\sqrt{3} \sqrt{r^2+1}+1}{2 \sqrt{r^2+2}}\right)$, The corresponding OR matrix is 
$$ O_{\KS1} = R[45^\circ,\ee_3]\,R[-\phi(r),\ve{\bar 1 }{ 1}{0}]\,R[-\theta(r), \ve{1}{1}{1}]$$
which yields the OR
\begin{equation*}
\no{1}{1}{1}_\gamma \parallel \no{0}{r}{1}\alp \aand \ve{1}{0}{\bar 1}_\gamma \parallel \ve{ 1}{1}{\bar r}\alp.
\end{equation*}
The application of $\PP$ yields the remaining $23$ \KS\ ORs (cf. Table~\ref{TableKSORA}). Note that, unlike Table~\ref{TableKSOR}, Table~\ref{TableKSORA} takes the tetragonality of the bct lattice into account and the bct vectors are given in a way that is consistent with the transformation strains and not up to crystallographic equivalence.
\begin{center}{\singlespacing
\begin{threeparttable}[ht] \captionsetup{format =plain}

\caption{The \KS\ orientation relationships. The corresponding trans\-formation strain in each row is given by 
$T_{\KS j}=R[\theta(r), P_{j}\ve{1}{1}{1}]\,R[\phi(r),P_j\ve{\bar 1 }{ 1}{0}]\, B_j$.}\label{TableKSORA}

 \begin{tabular}{cccccc}\hline
  OR\tnote{a} &  fcc plane\tnote{b} & bcc plane\tnote{c} & fcc direction\tnote{d} & bcc direction\tnote{e}& Bain Variant\tnote{f} \\
\hline    & & & & \vspace*{-1em} \\
 $\KS1 $&$ \no{1}{1}{1}_\gamma  $&$\no{0}{r}{1}\alp $&$\ve{1}{0}{\xo 1}_\gamma  $&$\ve{1}{1}{\xo r}\alp  $&$B_3$\\ 
 $\KS2 $&$ \no{\xo 1}{\xo 1}{\xo 1}_\gamma  $&$\no{\xo 1}{\xo r}{0}\alp $&$\ve{1}{0}{\xo 1}_\gamma  $&$\ve{r}{\xo 1}{\xo 1}\alp$ &$B_1$\\
 $\KS3 $&$ \no{1}{1}{1}_\gamma  $&$\no{1}{0}{r}\alp $&$\ve{ \xo 1}{1}{0}_\gamma  $&$\ve{\xo r}{1}{ 1}\alp $&$B_1 $\\
 $\KS4 $&$ \no{\xo 1}{\xo 1}{\xo 1}_\gamma  $&$\no{0}{\xo 1}{\xo r}\alp $&$\ve{ \xo 1}{1}{0}_\gamma  $&$\ve{\xo 1}{r}{\xo 1}\alp $&$B_2$\\ 
 $\KS5 $&$ \no{1}{1}{1}_\gamma  $&$\no{r}{1}{0}\alp $&$\ve{0}{\xo 1}{1}_\gamma  $&$\ve{1}{\xo r}{ 1}\alp  $&$B_2$\\
 $\KS6 $&$ \no{\xo 1}{\xo 1}{\xo 1}_\gamma  $&$\no{\xo r}{0}{\xo 1}\alp $&$\ve{0}{\xo 1}{1}_\gamma  $&$\ve{\xo 1}{\xo 1}{r}\alp$&$B_3 $\\
  &&&&\\
    $\KS7 $&$\no{\xo{1}}{1}{1}_\gamma  $&$\no{\xo 1}{r}{0}\alp $&$ \ve{1}{0}{1}_\gamma $&$\ve{r}{1}{ 1}\alp$&$B_1$\\
  $\KS8 $&$\no{1}{\xo 1}{\xo 1}_\gamma  $&$\no{0}{\xo r}{\xo 1}\alp $&$ \ve{1}{0}{1}_\gamma $&$\ve{1}{\xo 1}{r }\alp$&$B_3 $\\
    $\KS9 $&$ \no{\xo{1}}{1}{1}_\gamma  $&$\no{0}{1}{r}\alp $&$\ve{\xo 1}{\xo 1}{0}_\gamma $&$\ve{\xo 1}{\xo r}{ 1}\alp $&$B_2$\\
   $\KS10 $&$ \no{{1}}{\xo 1}{\xo 1}_\gamma  $&$\no{1}{0}{\xo r}\alp $&$\ve{\xo 1}{\xo 1}{0}_\gamma $&$\ve{\xo r}{\xo 1}{\xo 1}\alp$&$B_1$\\
  $\KS11 $&$ \no{\xo{1}}{1}{1}_\gamma  $&$\no{\xo r}{0}{1}\alp $&$ \ve{0}{1}{\xo 1}_\gamma $&$\ve{\xo 1}{1}{\xo r}\alp$&$B_3 $\\
  $\KS12 $&$ \no{{1}}{\xo 1}{\xo 1}_\gamma  $&$\no{r}{\xo 1}{0}\alp $&$\ve{0}{1}{\xo 1}_\gamma $&$\ve{1}{r}{\xo 1}\alp$&$B_2$\\
     &&&&\\ 
     $\KS13 $&$ \no{1}{\xo{1}}{1}_\gamma  $&$\no{1}{\xo r}{0}\alp $&$\ve{\xo 1}{0}{1}_\gamma $&$\ve{\xo r}{\xo 1}{ 1}\alp$&$B_1$\\
  $\KS14 $&$ \no{\xo 1}{{1}}{\xo 1}_\gamma  $&$\no{0}{r}{\xo 1}\alp $&$ \ve{\xo 1}{0}{1}_\gamma $&$\ve{\xo 1}{1}{r}\alp$&$B_3 $\\
  $\KS15 $&$\no{1}{\xo{1}}{{1}}_\gamma  $&$\no{0}{\xo 1}{r}\alp $&$\ve{1}{1}{0}_\gamma $&$\ve{1}{r}{ 1}\alp$&$B_2 $\\
    $\KS16 $&$\no{\xo 1}{{1}}{\xo {1}}_\gamma  $&$\no{\xo 1}{0}{r}\alp $&$ \ve{1}{1}{0}_\gamma $&$\ve{r}{1}{\xo 1}\alp$&$B_1$\\
   $\KS17 $&$ \no{1}{\xo{1}}{1}_\gamma  $&$\no{r}{0}{1}\alp $&$\ve{0}{\xo 1}{\xo 1}_\gamma $&$\ve{1}{\xo 1}{\xo r}\alp$&$B_3 $\\
  $\KS18 $&$ \no{\xo 1}{ 1}{\xo 1}_\gamma  $&$\no{\xo r}{1}{0}\alp $&$\ve{0}{\xo 1}{\xo 1}_\gamma $&$\ve{\xo 1}{\xo r}{\xo 1}\alp$&$B_2$\\
    &&&&\\     
 $\KS19 $&$ \no{1}{1}{\xo{1}}_\gamma  $&$\no{1}{r}{0}\alp $&$\ve{\xo 1}{0}{\xo 1}_\gamma  $&$\ve{\xo r}{1}{\xo 1}\alp $&$B_1$\\
 $\KS20 $&$ \no{\xo 1}{\xo 1}{{1}}_\gamma  $&$\no{0}{\xo r}{1}\alp $&$\ve{\xo 1}{0}{\xo 1}_\gamma $&$\ve{\xo 1}{\xo 1}{\xo r}\alp$&$B_3 $\\
  $\KS21 $&$ \no{1}{1}{\xo{1}}_\gamma  $&$\no{0}{1}{\xo r}\alp $&$\ve{1}{\xo 1}{0}_\gamma $&$\ve{1}{\xo r}{\xo 1}\alp$&$B_2$\\
  $\KS22 $&$ \no{\xo 1}{\xo 1}{{1}}_\gamma  $&$\no{\xo 1}{0}{r}\alp $&$\ve{1}{\xo 1}{0}_\gamma $&$\ve{r}{\xo 1}{ 1}\alp$&$B_1$\\
  $\KS23 $&$\no{1}{1}{\xo{1}}_\gamma  $&$\no{r}{0}{\xo 1}\alp $&$ \ve{0}{1}{1}_\gamma $&$\ve{1}{1}{r}\alp$&$B_3 $\\
  $\KS24 $&$\no{\xo 1}{\xo 1}{{1}}_\gamma  $&$\no{\xo r}{\xo 1}{0}\alp $&$ \ve{0}{1}{1}_\gamma $&$\ve{\xo 1}{r}{1}\alp$&$B_2 $\\
\hline

\end{tabular}

\begin{tablenotes}
\item[a] $\KS j$
\item[b] $P_{j}\no{1}{1}{1}_\gamma $
\item[c] $P_{j}\no{0}{r}{ 1 }\alp$
\item[d] $P_{j}\ve{1}{0}{\xo 1 }_\gamma $
\item[e] $P_{j}\ve{1}{1}{\xo r }\alp$
\item[f] $B_j=P_{j}B_3P_{j}^T$
\end{tablenotes}
\end{threeparttable}}
\end{center}

\subsection{Pitsch (PT)}
\noindent The transformation $T_{\PT1}$ is uniquely defined through our unified approach (cf. Section~\ref{SecUnif}) as the transformation that:
\begin{itemize}
 \item leaves the normal $\nn=\no{1}{1}{0}_\gamma$ and the direction $\vv=\ve{0}{0}{1}_\gamma$ unrotated,
 \item has pure stretch component $B_2$.
\end{itemize}
The resulting transformation strain is
\begin{equation*}
 T_{\PT1}=R[\psi(r), \ve{1}{0}{0}]\, B_2,
\end{equation*}
where $\psi(r) = - \arccos\left(\frac{\sqrt{2}+r}{\sqrt{2+r^2}}\right)$. The corresponding OR matrix is 
$$ O_{\PT1} = R[45^\circ,\ee_2]\,R[-\psi(r),\ve{1 }{ 0 }{0}]$$
which yields the OR
\begin{equation*}
\no{0}{1}{\bar 1}_\gamma \parallel \no{\bar{r}}{2}{\bar{r}}\alp \aand \ve{1}{0}{0}_\gamma \parallel \ve{1}{0}{\bar 1}\alp
\end{equation*}
The application of $\PP$ yields the remaining eleven \PT\ ORs (cf. Table~\ref{TablePOR}). 
\subsubsection*{Remark}
$O_{\PT1}$ also yields the parallelism $\ve{0}{1}{1}_\gamma \parallel \ve{1}{r}{1}\alp$ stated in \cite{Pitsch} (for $r=1$).

\begin{center}{\singlespacing
\begin{threeparttable}[ht] \captionsetup{format =plain}%
\caption{The Pitsch orientation relationships. The corresponding transformation strain in each row is given by 
$T_{\PT j}=R[\psi(r),P_{2j-1}\ve{1}{0}{0}]\,B_j$.}\label{TablePOR}

 \begin{tabular}{cccccc}\hline
  OR\tnote{a} &  fcc plane\tnote{b} & bcc plane\tnote{c} & fcc direction\tnote{d} & bcc direction\tnote{e}& Bain Variant\tnote{f} \\
\hline    & & & & \vspace*{-1em} \\
 \PT1 &  $\no{0 }{1 }{\xo 1 }_\gamma $ & $\no{\xo r }{2 }{\xo r }\alp$ & $\ve{1 }{0 }{0  }_\gamma $ & $\ve{1 }{0 }{\xo 1 }\alp$  & $B_2$\\ 
 \PT2 &  $\no{\xo 1 }{0 }{1 }_\gamma $ & $\no{\xo r }{\xo r }{2 }\alp$ & $\ve{ 0  }{1 }{0 }_\gamma $ & $\ve{\xo 1 }{1 }{0 }\alp$ &$B_3$ \\
 \PT3 &  $\no{1 }{\xo 1 }{0 }_\gamma $ & $\no{2 }{\xo r }{\xo r }\alp$ & $\ve{0 }{0 }{1 }_\gamma $ & $\ve{ 0}{\xo 1 }{1 }\alp$  &$B_ 1$\\
  & & & & \\
    \PT4 & $\no{1 }{1 }{0 }_\gamma $ & $\no{r }{2 }{\xo r }\alp$ &  $\ve{ 0}{ 0}{1 }_\gamma $& $\ve{1 }{0 }{1 }\alp$ & $B_2$\\
    \PT5 &  $\no{ 0}{\xo 1 }{1 }_\gamma $ & $\no{ r}{ \xo r}{2 }\alp$ & $\ve{ \xo 1}{ 0 }{0 }_\gamma $ & $\ve{\xo 1 }{\xo 1 }{0 }\alp$ &$B_3$\\
  \PT6 &  $\no{ \xo 1}{0 }{\xo 1}_\gamma $ & $\no{ \xo 2}{\xo r }{\xo r }\alp$ &  $\ve{ 0}{1 }{0 }_\gamma $& $\ve{ 0}{1 }{\xo 1 }\alp$&$B_1$ \\
     & & & & \\ 
     \PT7 &  $\no{\xo 1 }{\xo 1 }{0 }_\gamma $ & $\no{ \xo r}{\xo 2 }{\xo r }\alp$ & $\ve{0 }{0 }{1 }_\gamma $ &$ \ve{ \xo 1}{ 0}{ 1}\alp$& $B_2$\\
  \PT8 & $\no{ 0}{ 1}{1}_\gamma $ & $\no{\xo r }{r }{2 }\alp$ &$ \ve{ 1}{ 0}{0 }_\gamma $&$ \ve{ 1}{1 }{0 }\alp$&$B_3$ \\
   \PT9 &  $\no{1 }{0 }{\xo 1 }_\gamma $ & $\no{ 2}{r }{\xo r }\alp$ &$ \ve{ 0}{\xo 1 }{ 0 }_\gamma $ & $\ve{0}{\xo 1}{\xo 1}\alp$ &$B_1$ \\
    & & & & \\     
 \PT10 &  $\no{ \xo 1}{1 }{0 }_\gamma $ & $\no{\xo r }{2 }{r }\alp$ & $\ve{0  }{0 }{ \xo 1 }_\gamma $ &$ \ve{\xo 1 }{0 }{\xo 1 }\alp$ & $B_2$\\
  \PT11 &  $\no{0 }{\xo 1 }{ \xo 1}_\gamma $ & $\no{ \xo r}{\xo r }{\xo 2 }\alp$ & $\ve{1 }{0  }{0 }_\gamma $&$\ve{1 }{ \xo 1}{ 0}\alp$ &$B_3$\\
  \PT12 & $\no{1 }{ 0}{1}_\gamma $ & $\no{2 }{ \xo r}{ r}\alp$ & $ \ve{ 0}{1 }{0 }_\gamma $&$\ve{0 }{1 }{1 }\alp$&$B_1$ \\
\hline
\end{tabular}
\begin{tablenotes}
\item[a] $\PT_j$
\item[b] $P_{2j-1}\no{0}{1}{\xo 1}_\gamma$
\item[c] $P_{2j-1}\no{\xo r}{2}{\xo r}\alp$
\item[d] $P_{2j-1}\ve{1}{0}{0}_\gamma$
\item[e] $P_{2j-1}\no{1}{0}{\xo 1}\alp$
\item[f] $B_j=P_{2j-1}B_2P_{2j-1}^T$
\end{tablenotes}
\end{threeparttable}}
\end{center}

\subsection{Greninger-Troiano (GT)}
\noindent The transformation $T_{\GT1}$ is uniquely defined through our unified approach (cf. Section~\ref{SecUnif}) as the transformation that:
\begin{itemize}
 \item leaves the normal $\nn=\no{1}{1}{1}_\gamma$ and the direction $\vv=\ve{\bar{5}}{17}{\xo{12}}_\gamma$ unrotated,
 \item has pure stretch component $B_3$.
\end{itemize}
The resulting transformation strain is
\begin{equation*}
 T_{\GT1}=R[\xi(r), \ve{1}{1}{1}]\,R[\phi(r),\ve{ \bar 1 }{ 1 }{0}]\, B_3,
\end{equation*}
where $\xi(r) = \arccos\left(\frac{7^2+17^2 \sqrt{3}\sqrt{1 + r^2}}
 {\sqrt{2} \sqrt{5^2+12^2+17^2} \sqrt{7^2+17^2 + 17^2 r^2}}\right)$. The corresponding OR matrix is 
 $$ O_{\GT1} = R[45^\circ,\ee_3]\,R[-\phi(r),\ve{ \bar 1 }{ 1 }{0}]\,R[-\xi(r), \ve{1}{1}{1}]$$
 which yields the OR
\begin{equation*}
\no{1}{1}{1}_\gamma \parallel \no{0}{r}{1}\alp \aand \ve{\xo{12}}{\bar 5}{17}_\gamma \parallel \ve{\bar{7}}{\xo{17}}{17 r}\alp.
\end{equation*}
The application of $\PP$ yields the remaining $23$ \GT\ ORs (cf. Table~\ref{TableGTOR}). 
\subsubsection*{Example}
 Let $r=1.045$ (as in \cite{GT}) then $\no{1}{1}{1}: \no{0}{1}{1} \approx 1.26^\circ,   \ve{1}{1}{\bar{2}} : \ve{0}{1}{\bar 1} \approx2.82^\circ, \ve{1}{0}{\bar 1} : \ve{1}{1}{\bar 1} \approx2.94^\circ$ and $\ve{0}{\bar 1}{1} : \ve{1}{\bar 1}{1} \approx7.86^\circ.$

\begin{center}{\singlespacing
\begin{threeparttable}[ht] \captionsetup{format =plain}
\caption{The \GT\ orientation relationships. The corresponding transformation strain in each row is given by 
$T_{\GT j}=R[\xi(r), P_j \ve{1}{1}{1}]\,R[\phi(r),P_j\ve{\bar 1 }{ 1}{0}]\, B_j$.} \label{TableGTOR}

 \begin{tabular}{cccccc}\hline
  OR\tnote{a} &  fcc plane\tnote{b} & bcc plane\tnote{c} & fcc direction\tnote{d} & bcc direction\tnote{e}& Bain Variant\tnote{f} \\
\hline    & & & & \vspace*{-1em} \\
 $\GT1 $&$ \no{1}{1}{1}_\gamma  $&$\no{0}{r}{1}\alp $&$\ve{\xo{{12}}}{\xo  5}{17}_\gamma  $&$\ve{\xo  7}{\xo{17}}{{17r}}\alp  $&$B_3$\\ 
 $\GT2 $&$ \no{\xo 1}{\xo 1}{\xo 1}_\gamma  $&$\no{\xo 1}{\xo r}{0}\alp $&$\ve{\xo{17}}{5}{{{12}}}_\gamma  $&$\ve{\xo{17r}}{{17}}{7}\alp$ &$B_1$\\
 $\GT3 $&$ \no{1}{1}{1}_\gamma  $&$\no{1}{0}{r}\alp $&$\ve{17}{\xo{{12}}}{\xo  5}_\gamma  $&$\ve{{17r}}{\xo  7}{\xo{17}}\alp $&$B_1 $\\
 $\GT4 $&$ \no{\xo 1}{\xo 1}{\xo 1}_\gamma  $&$\no{0}{\xo 1}{\xo r}\alp $&$\ve{{{12}}}\xo{17}{5}_\gamma  $&$\ve{7}{\xo{17r}}{{17}}\alp $&$B_2$\\ 
 $\GT5 $&$ \no{1}{1}{1}_\gamma  $&$\no{r}{1}{0}\alp $&$\ve{\xo  5}{17}{\xo{{12}}}_\gamma  $&$\ve{\xo{17}}{{17r}}{\xo  7}\alp  $&$B_2$\\
 $\GT6 $&$ \no{\xo 1}{\xo 1}{\xo 1}_\gamma  $&$\no{\xo r}{0}{\xo 1}\alp $&$\ve{5}{{{12}}}{\xo{17}}_\gamma  $&$\ve{{17}}{ 7}{\xo{17r}}\alp$&$B_3 $\\
  &&&&\\
    $\GT7 $&$\no{\xo{1}}{1}{1}_\gamma  $&$\no{\xo  1}{r}{0}\alp $&$ \ve{\xo{17}}{\xo  5}{\xo{{12}}}_\gamma $&$\ve{\xo{{17r}}}{\xo{17}}{\xo  7}\alp$&$B_1$\\
  $\GT8 $&$\no{{1}}{\xo 1}{\xo 1}_\gamma  $&$\no{0}{\xo r}{\xo 1}\alp $&$ \ve{\xo{{12}}}{5}{\xo{17}}_\gamma $&$\ve{\xo 7}{{17}}{\xo{17r}}\alp$&$B_3 $\\
    $\GT9 $&$ \no{\xo{1}}{1}{1}_\gamma  $&$\no{0}{1}{r}\alp $&$\ve{{12}}{17}{\xo  5}_\gamma $&$\ve{7}{{17r}}{\xo{17}}\alp$&$B_2$\\
   $\GT10 $&$ \no{{1}}{\xo 1}{\xo 1}_\gamma  $&$\no{1}{0}{\xo r}\alp $&$\ve{{17}}{{{12}}}{5}_\gamma $&$\ve{{{17r}}}{7}{{17}}\alp$&$B_1$\\
  $\GT11 $&$ \no{\xo{1}}{1}{1}_\gamma  $&$\no{\xo  r}{0}{1}\alp $&$ \ve{\xo{5}}{{{12}}}{\xo{17}}_\gamma $&$\ve{17}{\xo  7}{{17r}}\alp$&$B_3 $\\
  $\GT{12} $&$ \no{{1}}{\xo 1}{\xo 1}_\gamma  $&$\no{r}{\xo 1}{0}\alp $&$\ve{5}{17}{\xo{{12}}}_\gamma $&$\ve{\xo{17}}{\xo{17r}}{7}\alp$&$B_2$\\
     &&&&\\ 
     $\GT13 $&$ \no{1}{\xo{1}}{1}_\gamma  $&$\no{1}{\xo  r}{0}\alp $&$\ve{17}{5}{\xo{{12}}}_\gamma $&$\ve{{17r}}{17}{\xo  7}\alp$&$B_1$\\
  $\GT14 $&$ \no{\xo 1}{{1}}{\xo 1}_\gamma  $&$\no{0}{r}{\xo 1}\alp $&$\ve{{{12}}}{\xo 5}{\xo{17}}_\gamma $&$\ve{7}{\xo{17}}{\xo{17r}}\alp$&$B_3 $\\
  $\GT15 $&$\no{1}{\xo{1}}{{1}}_\gamma  $&$\no{0}{\xo  1}{r}\alp $&$\ve{\xo{{12}}}{\xo{17}}{\xo{5}}_\gamma $&$\ve{\xo  7}{\xo{{17r}}}{\xo{17}}\alp$&$B_2 $\\
    $\GT16 $&$\no{\xo 1}{{1}}{\xo{1}}_\gamma  $&$\no{\xo 1}{0}{r}\alp $&$\ve{\xo{17}}{\xo{12}}{ 5}_\gamma $&$\ve{\xo{17r}}{\xo 7}{{17}}\alp$&$B_1$\\
   $\GT17 $&$ \no{1}{\xo{1}}{1}_\gamma  $&$\no{r}{0}{1}\alp $&$\ve{\xo  5}{{12}}{17}_\gamma $&$\ve{\xo{17}}{7}{{17r}}\alp$&$B_3 $\\
  $\GT18 $&$ \no{\xo 1}{1}{\xo 1}_\gamma  $&$\no{\xo r}{1}{0}\alp $&$\ve{5}{{17}}{{{12}}}_\gamma $&$\ve{{17}}{{{17r}}}{ 7}\alp$&$B_2$\\
    &&&&\\     
 $\GT19 $&$ \no{1}{1}{\xo{1}}_\gamma  $&$\no{1}{r}{0}\alp $&$\ve{17}{\xo  5}{{12}}_\gamma $&$\ve{{17r}}{\xo{17}}{7}\alp$&$B_1$\\
 $\GT20 $&$ \no{\xo 1}{\xo 1}{{1}}_\gamma  $&$\no{0}{\xo r}{1}\alp $&$\ve{{{12}}}{ 5}{{17}}_\gamma $&$\ve{7}{{17}}{{{17r}}}\alp$&$B_3 $\\
  $\GT21 $&$ \no{1}{1}{\xo{1}}_\gamma  $&$\no{0}{1}{\xo  r}\alp $&$\ve{\xo{{12}}}{17}{5}_\gamma $&$\ve{\xo  7}{{17r}}{17}\alp$&$B_2$\\
  $\GT22 $&$ \no{\xo 1}{\xo 1}{{1}}_\gamma  $&$\no{\xo 1}{0}{r}\alp $&$\ve{\xo{17}}{{{12}}}{\xo 5}_\gamma $&$\ve{\xo{17r}}{7}{\xo{17}}\alp$&$B_1$\\
  $\GT23 $&$\no{1}{1}{\xo{1}}_\gamma  $&$\no{r}{0}{\xo  1}\alp $&$\ve{\xo  5}{\xo{{12}}}{\xo{17}}_\gamma $&$\ve{\xo{17}}{\xo  7}{\xo{{17r}}}\alp$&$B_3 $\\
  $\GT24 $&$\no{\xo 1}{\xo 1}{{1}}_\gamma  $&$\no{\xo r}{\xo 1}{0}\alp $&$\ve{5}{\xo{17}}{\xo{12}}_\gamma $&$\ve{{17}}{\xo{17r}}{\xo 7}\alp$&$B_2 $\\
\hline

\end{tabular}

\begin{tablenotes}
\item[a] $\GT j$
\item[b] $P_{j}\no{1}{1}{1}_\gamma $
\item[c] $P_{j}\no{0}{r}{ 1 }\alp$
\item[d] $P_{j}\ve{\xo{12}}{\xo 5}{17 }_\gamma $
\item[e] $P_{j}\ve{\xo 7}{\xo{17}}{\xo{17r}}\alp$
\item[f] $B_j=P_{j}B_3P_{j}^T$
\end{tablenotes}
\end{threeparttable}}
\end{center}

\addtocontents{toc}{\protect\setcounter{tocdepth}{1}}
\subsection{Inverse Greninger-Troiano (GT')}
\noindent The transformation $T_{\GTp1}$ is uniquely defined through our unified approach (cf. Section~\ref{SecUnif}) as the transformation that:
\begin{itemize}
 \item leaves the normal $\nn=\no{\xo{17}}{\xo{7}}{{17}}_\gamma$ and the direction $\vv=\ve{1}{0}{1}_\gamma$ unrotated,
 \item has pure stretch component $B_3$.
\end{itemize}
The resulting transformation strain is
\begin{equation*}
 T_{\GTp1}=R[\iota(r), \ve{1}{0}{1}]\,R[-\psi(r),\ve{0}{1}{0}]\, B_3 = R[\iota(r), \ve{1}{0}{1}]\,R_{\PT2} B_3
\end{equation*}
where $\iota(r) = \arccos\left(\frac{17^2 \sqrt{2}\sqrt{
  2 + r^2} +  7^2 r }
 {\sqrt{17^2+17^2+7^2} \sqrt{2\cdot 17^2 + 7^2r^2 +17^2r^2}}\right)$. The corresponding OR matrix is 
 $$ O_{\GTp1} = R[45^\circ,\ve{0}{0}{1}]\,R[\psi(r),\ve{ 0}{ 1 }{0}]\,R[-\iota(r), \ve{1}{0}{1}]$$
 which yields the OR
\begin{equation*}
\no{\xo{17}}{\bar 7}{17}_\gamma \parallel \no{\bar{5r}}{\bar{12r}}{17}\alp \aand \ve{1}{0}{1}_\gamma \parallel \ve{1}{1}{r}\alp.
\end{equation*}
The application of $\PP$ yields the remaining $23$ \GTp\ ORs (cf. Table~\ref{TableGTpOR}). 
\begin{center}{\singlespacing
\begin{threeparttable}[ht] \captionsetup{format =plain}
\caption{The \GTp\ orientation relationships. The corresponding transformation strain in each row is given by 
$T_{\GTp j}=R[\iota(r), P_j\ve{1}{0}{1}]\,R[-\psi(r),P_j\ve{ 0 }{ 1}{0}]\, B_j$.}\label{TableGTpOR}

 \begin{tabular}{cccccc}\hline
  OR\tnote{a} &  fcc plane\tnote{b} & bcc plane\tnote{c} & fcc direction\tnote{d} & bcc direction\tnote{e}& Bain Variant\tnote{f} \\
\hline    & & & & \vspace*{-1em} \\
 $\GTp1 $&$ \no{\xo{17}}{\xo7}{17}_\gamma  $&$\no{\xo{5r}}{\xo{12r}}{17}\alp $&$\ve{1}{0}{1}_\gamma  $&$\ve{1}{1}{r}\alp  $&$B_3$\\ 
 $\GTp2 $&$ \no{\xo{17}}{7}{17}_\gamma  $&$\no{\xo{17}}{12r}{5r}\alp $&$\ve{\xo 1}{0}{\xo 1}_\gamma  $&$\ve{\xo r}{\xo 1}{\xo 1}\alp$ &$B_1$\\
 $\GTp3 $&$ \no{17}{\xo{17}}{\xo7}_\gamma  $&$\no{17}{\xo{5r}}{\xo{12r}}\alp $&$\ve{1}{1}{0}_\gamma  $&$\ve{r}{1}{1}\alp $&$B_1 $\\
 $\GTp4 $&$ \no{17}{\xo{17}}{7}_\gamma  $&$\no{5}{\xo{17r}}{12r}\alp $&$\ve{\xo 1}{\xo 1}{0}_\gamma  $&$\ve{\xo 1}{\xo r}{\xo 1}\alp $&$B_2$\\ 
 $\GTp5 $&$ \no{\xo7}{17}{\xo{17}}_\gamma  $&$\no{\xo{12r}}{17}{\xo{5r}}\alp $&$\ve{0}{1}{1}_\gamma  $&$\ve{1}{r}{1}\alp  $&$B_2$\\
 $\GTp6 $&$ \no{7}{17}{\xo{17}}_\gamma  $&$\no{12r}{5r}{\xo{17}}\alp $&$\ve{0}{\xo 1}{\xo 1}_\gamma  $&$\ve{\xo 1}{\xo 1}{\xo r}\alp$&$B_3 $\\
  &&&&\\
    $\GTp7 $&$\no{\xo{17}}{\xo 7}{\xo{17}}_\gamma  $&$\no{\xo{17}}{\xo{12r}}{\xo{5r}}\alp $&$ \ve{\xo1}{0}{1}_\gamma $&$\ve{\xo r}{1}{1}\alp$&$B_1$\\
  $\GTp8 $&$\no{\xo{17}}{7}{\xo{17}}_\gamma  $&$\no{\xo{5r}}{12r}{\xo{17}}\alp $&$ \ve{1}{0}{\xo 1}_\gamma $&$\ve{1}{\xo 1}{r}\alp$&$B_3 $\\
    $\GTp9 $&$ \no{17}{17}{\xo 7}_\gamma  $&$\no{5r}{17}{\xo{12r}}\alp $&$\ve{\xo 1}{1}{0}_\gamma $&$\ve{\xo1}{r}{1}\alp$&$B_2$\\
   $\GTp10 $&$ \no{17}{17}{7}_\gamma  $&$\no{17}{5r}{12r}\alp $&$\ve{1}{\xo 1}{0}_\gamma $&$\ve{r}{\xo 1}{\xo 1}\alp$&$B_1$\\
  $\GTp11 $&$ \no{7}{\xo{17}}{17}_\gamma  $&$\no{12r}{\xo{5r}}{17}\alp $&$ \ve{0}{1}{1}_\gamma $&$\ve{\xo1}{1}{r}\alp$&$B_3 $\\
  $\GTp12 $&$ \no{\xo{7}}{\xo{17}}{17}_\gamma  $&$\no{\xo{12r}}{\xo{17}}{5r}\alp $&$\ve{0}{\xo 1}{\xo 1}_\gamma $&$\ve{1}{\xo r}{\xo 1}\alp$&$B_2$\\
     &&&&\\ 
     $\GTp13 $&$ \no{17}{7}{\xo{17}}_\gamma  $&$\no{17}{12r}{\xo{5r}}\alp $&$\ve{1}{0}{1}_\gamma $&$\ve{r}{\xo 1}{1}\alp$&$B_1$\\
  $\GTp14 $&$ \no{17}{\xo7}{\xo{17}}_\gamma  $&$\no{5r}{\xo{12r}}{\xo{17}}\alp $&$\ve{\xo 1}{0}{\xo 1}_\gamma $&$\ve{\xo 1}{1}{\xo r}\alp$&$B_3 $\\
  $\GTp15 $&$\no{\xo{17}}{\xo{17}}{\xo7}_\gamma  $&$\no{\xo{5r}}{\xo{17}}{\xo{12r}}\alp $&$\ve{1}{\xo 1}{0}_\gamma $&$\ve{1}{\xo r}{1}\alp$&$B_2 $\\
    $\GTp16 $&$\no{\xo{17}}{\xo{17}}{7}_\gamma  $&$\no{\xo{17}}{\xo{5r}}{12r}\alp $&$\ve{\xo 1}{1}{0}_\gamma $&$\ve{\xo r}{1}{\xo 1}\alp$&$B_1$\\
   $\GTp17 $&$ \no{\xo 7}{17}{17}_\gamma  $&$\no{\xo{12r}}{5r}{17}\alp $&$\ve{0}{\xo 1}{1}_\gamma $&$\ve{1}{\xo 1}{r}\alp$&$B_3 $\\
  $\GTp18 $&$ \no{7}{17}{17}_\gamma  $&$\no{12r}{17}{5r}\alp $&$\ve{0}{ 1}{\xo 1}_\gamma $&$\ve{\xo 1}{r}{\xo 1}\alp$&$B_2$\\
    &&&&\\     
 $\GTp19 $&$ \no{17}{\xo 7}{17}_\gamma  $&$\no{17}{\xo{12r}}{5r}\alp $&$\ve{\xo 1}{0}{ 1}_\gamma $&$\ve{r}{1}{\xo{1}}\alp$&$B_1$\\
 $\GTp20 $&$ \no{17}{7}{17}_\gamma  $&$\no{5r}{12r}{17}\alp $&$\ve{1}{0}{\xo 1}_\gamma $&$\ve{\xo 1}{\xo 1}{r}\alp$&$B_3 $\\
  $\GTp21 $&$ \no{\xo{17}}{17}{7}_\gamma  $&$\no{\xo{5r}}{17}{12r}\alp $&$\ve{\xo 1}{\xo 1}{0}_\gamma $&$\ve{1}{r}{\xo{1}}\alp$&$B_2$\\
  $\GTp22 $&$ \no{\xo{17}}{17}{\xo 7}_\gamma  $&$\no{\xo{17}}{5r}{\xo{12r}}\alp $&$\ve{1}{1}{0}_\gamma $&$\ve{\xo r}{\xo 1}{1}\alp$&$B_1$\\
  $\GTp23 $&$\no{\xo 7}{\xo{17}}{\xo{17}}_\gamma  $&$\no{\xo{12r}}{\xo{ 5r}}{\xo{17}}\alp $&$\ve{0}{\xo 1}{ 1}_\gamma $&$\ve{1}{1}{\xo{r}}\alp$&$B_3 $\\
  $\GTp24 $&$\no{7}{\xo{17}}{\xo{17}}_\gamma  $&$\no{12r}{\xo{17}}{\xo{ 5r}}\alp $&$\ve{0}{1}{\xo 1}_\gamma $&$\ve{\xo 1}{\xo 1}{1}\alp$&$B_2 $\\
\hline

\end{tabular}

\begin{tablenotes}
\item[a] $\GTp j$
\item[b] $P_{j}\no{\xo{17}}{\xo{7}}{17}_\gamma $
\item[c] $P_{j}\no{\xo{5r}}{\xo{12r}}{ 17 }\alp$
\item[d] $P_{j}\ve{1}{0}{1}_\gamma $
\item[e] $P_{j}\ve{1}{1}{r}\alp$
\item[f] $B_j=P_{j}B_3P_{j}^T$
\end{tablenotes}
\end{threeparttable}}
\end{center}
\newpage
\section{The group $\PP$}\label{App}
The elements of $\mathcal{P}^{24}$ in the standard basis $\lbrace \ee_1,\ee_2,\ee_3\rbrace$ are given by
\small
{\singlespacing
\begin{align*}
  P_1 = \mathbf{1} = \begin{pmatrix} 1 & 0 & 0 \\ 0 & 1 & 0\\ 0 & 0 & 1 \end{pmatrix}&,\quad &
  P_{2} = R[180^\circ, \ee_1-\ee_3]=\begin{pmatrix} 0 & 0 & -1 \\ 0 & -1 & 0\\ -1 & 0 & 0 \end{pmatrix},& \\ 
  P_3 = R[120^\circ, \ee_1+\ee_2+\ee_3]=\begin{pmatrix} 0 & 0 & 1 \\ 1 & 0 & 0\\ 0 & 1 & 0 \end{pmatrix}&, \quad &
   P_{4} = R[180^\circ, \ee_2-\ee_3]=\begin{pmatrix} -1 & 0 & 0 \\ 0 & 0 & -1\\ 0 & -1 & 0 \end{pmatrix},&\\
     P_5 = R[-120^\circ, \ee_1+\ee_2+\ee_3]=\begin{pmatrix} 0 & 1 & 0 \\ 0 & 0 & 1\\ 1 & 0 & 0 \end{pmatrix}&, \quad &   
   P_{6} = R[180^\circ, \ee_1-\ee_2]=\begin{pmatrix} 0 & -1 & 0 \\ -1 & 0 & 0\\ 0 & 0 & -1 \end{pmatrix},&\\
     P_{7} = R[-90^\circ, \ee_2]=\begin{pmatrix} 0 & 0 & -1 \\ 0 & 1 & 0\\ 1 & 0 & 0 \end{pmatrix}&, \quad & 
         P_{8} = R[180^\circ, \ee_1]=\begin{pmatrix} 1 & 0 & 0 \\ 0 & -1 & 0\\ 0 & 0 & -1 \end{pmatrix},& \\ 
     P_{9} = R[180^\circ, \ee_2+\ee_3]=\begin{pmatrix} -1 & 0 & 0 \\ 0 & 0 & 1\\ 0 & 1 & 0 \end{pmatrix}&, \quad &   
      P_{10} = R[-120^\circ, \ee_1-\ee_2+\ee_3]=\begin{pmatrix} 0 & 0 & 1 \\ -1 & 0 & 0\\ 0 & -1 & 0 \end{pmatrix}&, \\
      P_{11} = R[90^\circ, \ee_3]=\begin{pmatrix} 0 & -1 & 0 \\ 1 & 0 & 0\\ 0 & 0 & 1 \end{pmatrix}&, \quad &
      P_{12} = R[120^\circ, \ee_1+\ee_2-\ee_3]=\begin{pmatrix} 0 & 1 & 0 \\ 0 & 0 & -1\\ -1 & 0 & 0 \end{pmatrix},& \\
       P_{13} = R[180^\circ, \ee_1+\ee_3]=\begin{pmatrix} 0 & 0 & 1 \\ 0 & -1 & 0\\ 1 & 0 & 0 \end{pmatrix}&, \quad &      
         P_{14} = R[180^\circ, \ee_2]=\begin{pmatrix} -1 & 0 & 0 \\ 0 & 1 & 0\\ 0 & 0 & -1 \end{pmatrix},& \\ 
          P_{15} = R[90^\circ, \ee_1]=\begin{pmatrix} 1 & 0 & 0 \\ 0 & 0 & -1\\ 0 & 1 & 0 \end{pmatrix}&, \quad &    
           P_{16} = R[-120^\circ, \ee_1+\ee_2-\ee_3]=\begin{pmatrix} 0 & 0 & -1 \\ 1 & 0 & 0\\ 0 & -1 & 0 \end{pmatrix},& \\       
 P_{17} = R[-90^\circ, \ee_3]=\begin{pmatrix} 0 & 1 & 0 \\ -1 & 0 & 0\\ 0 & 0 & 1 \end{pmatrix}&, \quad &         
    P_{18} = R[120^\circ, -\ee_1+\ee_2+\ee_3]=\begin{pmatrix} 0 & -1 & 0 \\ 0 & 0 & 1\\ -1 & 0 & 0 \end{pmatrix},&  \\ 
     P_{19} = R[90^\circ, \ee_2]=\begin{pmatrix} 0 & 0 & 1 \\ 0 & 1 & 0\\ -1 & 0 & 0 \end{pmatrix}&, \quad & 
   P_{20} = R[180^\circ, \ee_3]=\begin{pmatrix} -1 & 0 & 0 \\ 0 & -1 & 0\\ 0 & 0 & 1 \end{pmatrix},& \\
      P_{21} = R[-90^\circ, \ee_1]=\begin{pmatrix} 1 & 0 & 0 \\ 0 & 0 & 1\\ 0 & -1 & 0 \end{pmatrix}&, \quad &         
      P_{22} = R[-120^\circ, -\ee_1+\ee_2+\ee_3]=\begin{pmatrix} 0 & 0 & -1 \\ -1 & 0 & 0\\ 0 & 1 & 0 \end{pmatrix},&  \\       
       P_{23} = R[180^\circ, \ee_1+\ee_2]=\begin{pmatrix} 0 & 1 & 0 \\ 1 & 0 & 0\\ 0 & 0 & -1 \end{pmatrix}&, \quad & 
      P_{24} = R[120^\circ, \ee_1-\ee_2+\ee_3]=\begin{pmatrix} 0 & -1 & 0 \\ 0 & 0 & -1\\ 1 & 0 & 0 \end{pmatrix}.& 
\end{align*}}
\normalsize

\end{document}